\documentclass[%
 reprint,
 amsmath,amssymb,
 aps,
 prm,
floatfix,
]{revtex4-2}

\usepackage{graphicx}%
\usepackage{dcolumn}%
\usepackage{bm}%
\usepackage{mhchem}
\usepackage{xcolor}
\usepackage{comment}

\newcommand{\gaasx}{$\text{Ga}_x\text{As}_{1-x}~$}
\newcommand{\SI}{S.M.\cite{supplmat}}
\begin{document}
\newcommand{\rev}[1]{#1}

\preprint{APS/123-QED}

\title{Modeling the Ga/As binary system across temperatures \\and compositions from first principles}

\author{Giulio Imbalzano}
\author{Michele Ceriotti}
\affiliation{%
 Laboratory of Computational Science and Modeling, IMX, \'Ecole Polytechnique F\'ed\'erale de Lausanne, 1015 Lausanne, Switzerland
}%

\date{\today}%

\begin{abstract}
Materials composed of elements from the third and fifth columns of the periodic table display a very rich behavior, with the phase diagram usually containing a metallic liquid phase and a polar semiconducting solid. 
As a consequence, it is very hard to achieve transferable empirical models of interactions between the atoms that can reliably predict their behavior across the temperature and composition range that is relevant to the study of the synthesis and properties of III/V nanostructures and devices. 
We present a machine-learning potential trained on density functional theory reference data that provides a general-purpose model for the \gaasx system. 
We provide a series of stringent tests that showcase the accuracy of the potential, and its applicability across the whole binary phase space, computing with ab initio accuracy a large number of finite-temperature properties as well as the location of phase boundaries. We also show how a committe model can be used to reliably determine the uncertainty induced by the limitations of the ML model on its predictions, to identify regions of phase space that are predicted with insufficient accuracy, and to iteratively refine the training set to achieve consistent, reliable modeling. 
\end{abstract}

\maketitle

\section{\label{sec:intro}Introduction}

Gallium arsenide, the prototypical III-V semiconductor, has excellent electronic and optical properties, which have been thoroughly studied from both an experimental and theoretical point of view\cite{blakemore1982semiconducting,senichev2018electronic,han2013tunable,woggon1997optical,ohno1999properties, barrigon2019synthesis}. Recent technological advances in the synthesis of GaAs have allowed to move towards the miniaturization of devices\cite{friedl2020remote,barrigon2019synthesis,duan2003high,mcintyre2020semiconductor}, which in turn has opened the doors to the fine-tuning of their electronic properties for many applications, such as light and x-ray detectors\cite{wallentin2014hard,boulanger2016characterization,chen2018analysis}, lasers\cite{kim2017monolithic}, and topologically-protected qubits\cite{friedl2018template}.

From the theoretical side, molecular dynamics has been widely used to understand the physics behind experimental observations \cite{chrobak2007nondislocation}, the behaviour of the material under ionizing radiation \cite{bjorkas2006damage}, the growth of semiconductor thin films\cite{drautz2007analytic}, and, more recently, the self-catalyzed growth of GaAs nanowires\cite{oliveira2020interatomic}. However, each one of these studies required the use of \textit{ad hoc} potentials that are tied to a specific region of the binary phase space. A general and transferable potential would not only allow to study complex phenomena involving multiple phases, but also eliminate the need to train one potential from scratch for each novel application that arises due to technological progress. A more modern approach based on the use of machine learning has created the opportunity to generate potentials that can cover the whole phase space, while retaining the accuracy of \textit{ab initio} methods at roughly the cost of an empirical potential\cite{behler2007generalized,bartok2010gaussian,zuo2020performance,shapeev2016moment}. This has already proven to be an invaluable tool to study at an unprecedented level of computational accuracy systems that are otherwise inaccessible to computational investigation\cite{sosso2013fast,zamani20203d,morawietz2016how}.

Although the typical accuracy achievable with these machine learning potentials (MLP) is known, it is important to carefully assess the potentials that we use. For most purposes, it is enough to train the MLP only on a selected region of the phase space to investigate a particular phenomenon of interest. However, to have a stable, accurate, and fully transferable model, one needs to include a much more varied set of structures in the training. Examples of potentials that try to model multiple regions of the phase space include, but are not limited to, iron\cite{dragoni2018achieving}, silicon\cite{bartok2018machine}, GeTe\cite{sosso2012neural}, $\text{Mn}_x\text{Ge}_y$\cite{mangold2020transferability}, and Ga\cite{niu2020ab}. This approach clearly introduces some challenges regarding both the limits of the description used to represent the structures and the optimal construction of a large dataset.
For the case of GaAs, and in general III-V semiconductors, the electronic behaviour of the system changes greatly when moving across the binary phase diagram, which contains both metallic and semiconducting phases. A potential for the study of the phase diagram of pure gallium has been recently proposed, providing an accurate picture of the phase transitions of this material\cite{niu2020ab}, but it does not explore the complications deriving from the addition of a second element, nor it provides a full survey of the properties at finite temperature.

In this work, we discuss the construction and extensively demonstrate the application of a machine learning potential that can be used to study the \gaasx system in a wide range of temperatures and pressures at any stoichiometry. To train the potential, we rely on the method proposed by Behler and Parrinello\cite{behler2007generalized} as implemented by Singraeber \textit{et al.}\cite{singraber2019library}. Therefore, we describe the local environments in terms of ``symmetry functions'' and use the flexibility of a neural network to parametrize the interactions among the atoms, following the successful example of other similar works\cite{sosso2012neural,sosso2013fast,morawietz2016how,quaranta2017proton,artrith2016implementation}. In order to give a sense of the accuracy and reliability of the potential, we thoroughly investigate the binary phase diagram and compute various properties for both the solid and liquid phases at finite temperature. For the solid phase, we present the results for the heat capacity and thermal expansion from cryogenic temperatures up to melting. For the liquid, we show the density, diffusion, and radial pair distribution functions of Ga, As, and GaAs. We also present encouraging result for the transferability of the potential to the study of amorphous GaAs. At last, we conclude with the binary phase diagram predicted by our potential.

As an additional benefit, one of the strengths of this type of potentials is the knowledge that they can be incrementally improved in the future, should the need arise. On one side, should the potential be found to be unreliable in a particular region of the phase space, one can increase its accuracy in this region by adding new configurations to the dataset. On the other side, if the level of theory of the DFT reference were to be found inadequate, one can recompute all, or a subset, of the configurations at a higher level of theory, thus improving the results while retaining the low computational cost.

The paper is divided into three sections. In the first part we provide the details of the generation of the potential, outlining both the architecture that has been employed and the database generation. In the second part we validate the potential, demonstrating the quality of the fit while comparing it to existing literature. In the third section we show the results of simulations and calculations beyond those used for training. This is intended as a test of the transferability of the potential, while also assessing the level of theory chosen in this work. Finally, we draw the conclusions on the current state of the potential and the possibility of using it to model systems of particular interest.

\section{\label{sec:methods}Methods}

\subsection{Architecture of the potential}

While the choice of the representation and the machine learning algorithm in use are important for the quality of the fitting, recent works have shown that different frameworks tend to achieve similar performance for a given dataset\cite{nguyen2018comparison,zuo2020performance}. 

\subsubsection{Details of the MLP}

For our potential, we have decided to follow the work done by Behler and Parrinello\cite{behler2007generalized}, since it has already been thoroughly tested on a number of different materials and it has been shown to perform well on systems similar to ours. We use the implementation by Singraber and Dellago\cite{singraber2019parallel}. Transforming this potential into a different one would be as easy as refitting the training set that we have generated using a different package.%

Since the symmetry functions (SF) are tied to the system\cite{behler2011atom} and tend to be very correlated to each other\cite{goscinski2021role}, we use an unsupervised method\cite{imbalzano2018automatic} based on the CUR decomposition\cite{mahoney2009cur} to select a small set of uncorrelated SFs out of a larger pool, generated to span the possible space of interactions. 

For this system, we find that a set of 64 SFs for each species (thus a total of 128 SFs) is sufficient to describe the local environments. The selection has been repeated after the addition of every set of new training structures, to ensure that the selection is able to capture all the novel relevant correlations. We observe that late additions to the training set have little effect on the choice of the SFs, indicating the robustness of our method. Details of the selection and the SFs that have been ultimately used for the potential can be found in the files provided for the potential.

The regression scheme that we use in this work is a feed-forward neural network with 2 layers and 24 nodes per layer, for a total of 4370 parameters, 2185 for each species, that must be optimized. The optimization procedure is carried out minimizing the errors between the predicted energies and forces with respect to the known DFT values, using a parallel Kalman filter implementation\cite{singraber2019parallel}.

\subsubsection{Uncertainty estimation}
The ability to quantify the uncertainty deriving from the use of a machine learning model is very important, since it can be used to assess the confidence of results and, possibly, to indicate configurations that are beyond the phase space which is well represented in the training database. In the case of neural network potentials the uncertainty quantification can be achieved by training multiple models, and using for each configuration $A$ their mean $\bar{y}(A)=\frac{1}{M}\sum_m y^{(m)}(A)$ as the committee prediction, and their standard deviation $\sigma(A)$ as a measure of the uncertainty\cite{musil2019fast,novikov2020mlip,rossi2020simulating,anelli2018generalized,anelli2020learning,schran2020committee}.

If we use this approach, we have to account for the limited dataset and implicit correlation of the models with a correction parameter $\alpha$\cite{musil2019fast,imbalzano2020uncertainty} that ensures that the predicted variance matches the expected variance over an unrelated validation set, as 
\begin{equation}\label{eq:alpha}
    \alpha^2 =-\frac{1}{M} + \frac{M-3}{M-1}  \frac{1}{N}\sum_n^N \frac{(y_n - \bar{y}(A_n)^2)}{\sigma(A_n)^2}.
\end{equation}
All the calculations in this paper have been performed using an ensemble of $M=4$ potentials independently trained on different (but overlapping) subsets of the same training set and starting from different initial weights. The average of the forces and energies is used to drive the dynamics and provide numerical estimates of the confidence intervals for some properties. The set of structures used to estimate $\alpha$ has been removed from the full dataset before starting the training procedure. More details about the dataset generation are presented in section \ref{sec:database}

For properties computed on single structures (e.g. elastic moduli), this method allows to distinguish low accuracy and low precision predictions. During molecular dynamics simulations, the uncertainty computed in this way allows to identify configurations that are poorly predicted, suggesting that the simulation is moving in previously unexplored regions of the phase space. 
It has recently been shown that this kind of uncertainty estimation can also be used to compute the error on thermodynamical averages, that results from the ML approximation of the reference potential\cite{imbalzano2020uncertainty}. We show an early application of this approach to the evaluation of the ML error for the pair distribution functions and the melting point of Ga, As, and GaAs. 

On top of the uncertainty deriving from the use of a ML model, we also take into account the statistical error due to the finite time of the MD simulations, computed using the block averaging method.

\subsection{Database generation and details\label{sec:database}}

To generate a potential able to cover the full binary phase diagram, it is necessary to add training structures of all the various phases of GaAs, Ga, As and their relative interfaces.
We use concepts that have already appeared in the literature to create a database that spans all of the phase space of interest.\cite{deri+20ncomm,bartok2018machine} The structures that are contained in the database have been obtained using three related but different approaches. 
We start with a potential limited to a small part of the phase space, we extend it to reproduce static properties of all of the phases of interest and we finally ensure its stability by using an active learning-like procedure on more challenging simulations. 
\rev{The final dataset is composed of 1921 structures, belonging to crystal structures that are stable around standard conditions, liquid phases and various interfaces (solid-liquid, liquid-vacuum, and solid-vacuum). We note that other crystal structures, as well as amorphous and disordered phases were not explicitly included and our potential should be used with caution to simulate these specific systems. 
We demonstrate the transferability of our potential outside the phases that are included explicitly in Fig.~ \ref{fig:RDF_aGaAs}, where we compute  radial pair distribution functions of amorphous GaAs, in excellent agreement with experiments. For these cases, the availability of a predicted uncertainty is essential to determine the level of reliability of the corresponding predictions.}

\subsubsection{A potential for the interface}
As a first iteration we use a potential generated to model the interface between liquid Ga and crystalline GaAs (in either WZ or ZB phase), that was used to study the differences in the ordering between the two polar surfaces in the [111] growth direction\cite{zamani20203d}. 
Given the limited scope of this potential, the training set was generated from short \textit{ab initio} MD simulations of the interface, followed by more extensive sampling that combines a preliminary MLP and a DFT correction, as discussed in Ref.~\citenum{zamani20203d}

\subsubsection{Complementing the potential}
We extend this potential by explicitly including structures needed to compute known static properties, such as lattice constants, elastic constants, surface decohesion energies, surface reconstructions, point defects and selected plane defects for Ga, As, and GaAs.

For this purpose, we have generated the structures needed to compute these properties, either as single-point calculations (e.g. lattice constants) or by relaxing the structure (e.g. defects and surfaces), thus obtaining a sequence of correlated structures. From the relaxations we have chosen to keep for training only a few out of all of the generated structures, making sure to include the initial, the final and some intermediate steps whose energies are found to be significantly different compared to the initial and final configurations. 
Adding the discarded configurations to the training set would have an impact only on the training time, but not on the computational cost of the MLP in production. However, we prefer to keep a smaller and more efficient training set in order to reduce the future cost of recomputing the structures at a different level of theory.
This additional set of 557 structures yields a potential that is able to correctly reproduce these static properties across all these phases, but does not guarantee stability at high temperature or at intermediate stoichiometries.

\subsubsection{Iterating over uncertain configurations}
To complete the potential and ensure that it is reliable for all of the properties that we want to model, we have used an offline active learning strategy, introducing in the dataset some of the structures generated throughout the validation process. Whenever we observe an uncertainty in the committee higher than a threshold (arbitrarily chosen to be 5 times the RMSE) during a simulation, we gather the structures that are poorly predicted and select a small and representative set of configurations for retraining. The structures are chosen either by using a farthest point sampling strategy, i.e. spanning as uniformly as possible the space of configurations, or by iteratively adding those with the highest uncertainty, stopping when the predictions become accurate.

We observe that with this procedure we have added many structures of liquid \gaasx with $0.05 < x < 0.45$ and $0.55 < x < 0.95$, which were initially found to be poorly predicted. This is an obvious consequence of the previous training procedures, where stoichiometries of $x= {0,0.5,1}$ were favoured, leaving the other regions of the phase space poorly sampled. Similarly, we add a number of structures of liquid Ga at high pressure, a region that we had not initially included in the training but that it is of great technological relevance.

\subsubsection{Details of the DFT calculations}
All the DFT calculations have been run using \textsc{Quantum ESPRESSO}\cite{giannozzi2009quantum}. In order to ensure an absolute convergence of the calculation to below 1 meV/atom we have used an energy cut-off of 50 Ry and a density of 6.5 k-points \AA. The GGA approximation with PBE exhange-correlation function\cite{perdew1996generalized} has been used, together with ultrasoft pseudopotentials\cite{vanderbilt1990soft} from the SSSP accuracy library (version 0.7)\cite{prandini2018precision}.

In order to minimize the errors arising from minute differences in the k-point grid, we have used, as consistently as possible, similarly sized supercells, with average dimensions of 12x14x40 \AA. The elongated shape and large size have been chosen in order to accommodate two different bulk systems and their interface in a single supercell (e.g. the interface between liquid \gaasx and solid GaAs from the original dataset). This also helped to ensure that the cell was large enough to avoid interactions among periodic replicas for defect calculations.

\subsection{Molecular dynamics\label{sec:MD}}
To test the potential beyond the properties that can be probed with single-point calculations, we run MD simulations for the system in its solid and liquid form, together with various interfaces. Since our investigation includes the evaluation of these properties at very low temperature, it is necessary to explicitly include the effects of the quantum motion of the nuclei to recover the correct properties.

Path integral molecular dynamics (PIMD) is a formalism needed to include nuclear quantum effects (NQE) into the simulation, which relies on the isomorphism between a nucleus and a chain of \textit{P} beads connected by springs, where \textit{P} must be increased to ensure convergence to the quantum Boltzmann distribution. More details on the theory of PIMD can be found elsewhere\cite{ceriotti2016nuclear,markland2018nuclear}, whereas from our perspective it is important to mention that simulating a system of \textit{P} beads has the same computational cost of running \textit{P} parallel simulations of the same system. 

All the MD and PIMD simulations have been run using i-PI\cite{kapil2019ipi} to propagate the dynamics and LAMMPS\cite{plimpton1995fast} with the n2p2 plugin\cite{singraber2019library} to compute energies and forces at every step. Boxes of about 300 atoms have been used in most cases for determining the properties, unless specified. 
The temperature has been constrained using a combination of a generalized Langevin\cite{ceriotti2010colored} and stochastic velocity rescaling thermostats\cite{bussi2007canonical}, whereas the pressures, when needed, have been constrained using an isotropic Bussi-Zykova-Parrinello barostat\cite{bussi2009isothermal} as implemented in i-PI. Solid/liquid coexistence simulations were performed with an anisotropic Np$_z$T scheme. 
A timestep of 4 fs has been used to integrate the equations of motion. %

\section{\label{sec:validation}Validating the potential}

The final database generated as explained in Sec. \ref{sec:database} is composed of 1921 structures, out of which 100 have been excluded from the training procedure and have been used both as a final test set and to compute the $\alpha$ parameter from eq. \ref{eq:alpha}. Each potential is trained on the remaining 1821 structures, 20\% of which, randomly chosen for each potential, are used for internal validation. 

\begin{figure}[tbp]
  \centering
  \includegraphics[width=\columnwidth]{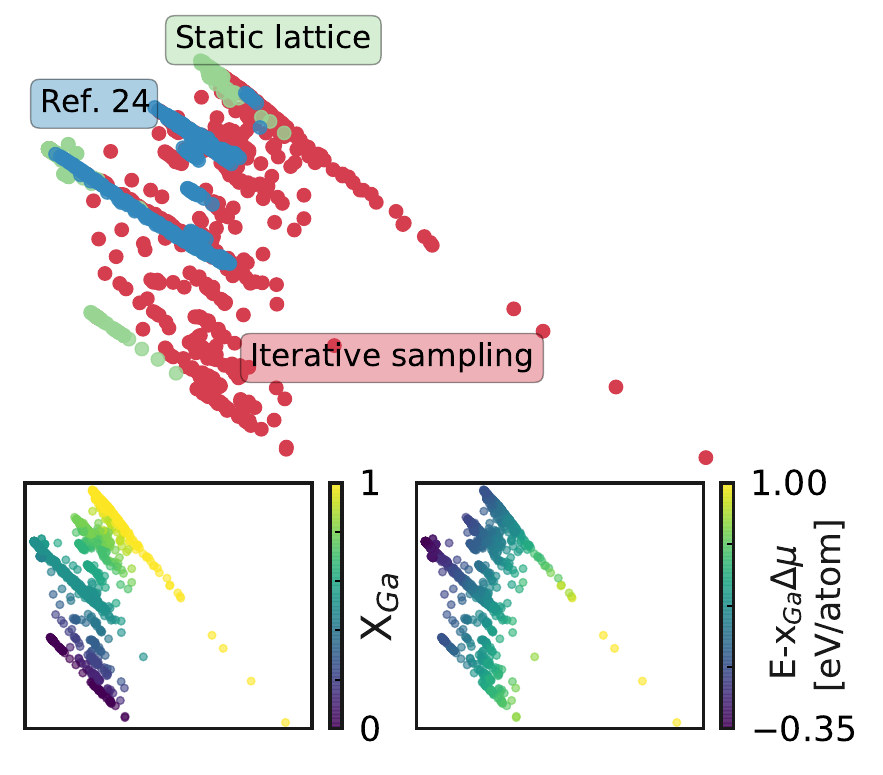}
  \caption{KPCoVR map\cite{helf+20mlst} of the configurations used to fit the final NNP. The map uses an equal mix of PCA and linear regression of the energy relative to the trivial combination $xE_{Ga} + (1-x)E_{As}$ 
  ($\alpha = 0.5$, following the convention of \citenum{helf+20mlst}) to illustrate the similarity among the structures. Different colours highlight the origin of the data, as presented in section \ref{sec:database}.
  The subplots present the same map, coloured according to the stoichiometry (left) and the hull distance (right), the same quantity used for KPCovR construction underlying the map.
  }
  \label{fig:kpcovrmap}
\end{figure}

Figure \ref{fig:kpcovrmap} illustrates the similarity among the structures that are present in the database. The colours represent the origin of the structures, following the methods detailed in section \ref{sec:database}. The layout of the points has been obtained with a KPCovR projection\cite{helf+20mlst} and reflects the composition and stability of different configurations. It can be noticed that the initial configurations are limited to a small region at very precise stoichiometries and low relative energy, while the iterative sampling allows to fill the gaps between the regions and to incorporate defective, high energy structures. 

\begin{figure}[tbph]
  \centering
  \includegraphics[width=0.9\columnwidth]{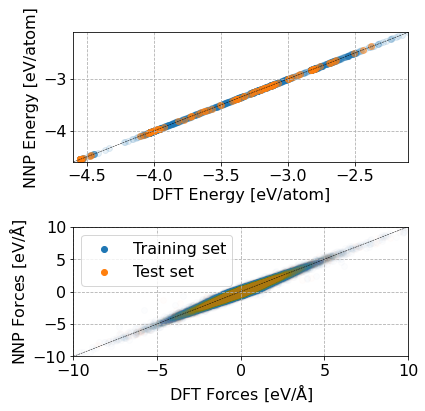}
  \caption{Parity plot comparing the energies (top) and forces (bottom) predicted with the NNP against the reference values from DFT. The test set is an independent set of 100 structures which have been excluded from the training procedure. The dashed line is added as a guide for the perfect match between prediction and reference.}
  \label{fig:parity_all}
\end{figure}

Figure \ref{fig:parity_all} shows the parity plots for energies (top) and forces (bottom). In these plots, we refer to ``training set'' to indicate the full 1821 structures that have been used for training, even if not all of them appear in every potential, and the ``test set'' is the set of 100 structures initially removed from the database. The RMSE for the committee computed on the test set is found to be 2.4 meV/atom for the energies and 109 meV/\AA~for the forces. We correct for the intrinsic correlation in the dataset with a factor $\alpha = 2.2$. These values show a very accurate fitting, particularly when one considers the very diverse set of structures used in the training.

While these values provide a sense of the typical error for this potential, they do not necessarily reflect the ultimate accuracy when computing specific, physically and technologically relevant observables. To provide a compelling demonstration of the versatility and limits of the potential, we compute a selection of properties and compare them to DFT calculations or experimental values. 
The static lattice properties that we compute are closely related to structures that are part of the training set, and so they do not fully report on the transferability of the model but rather on the quality of the fit. Results on finite-temperature properties, discussed in section \ref{sec:results}, provide a complementary perspective on the behavior of the ML potential and the underlying DFT reference.

We also provide the results obtained for the same properties with two of the most successful empirical potentials that have been published in the past for GaAs, and are fitted to experimental data. The first is the so called ANNK potential, from the initials of the authors, which has the form of a modified Tersoff potential\cite{albe2002modeling} and has seen wide use for the study of the effect of radiation on crystalline GaAs. The second is the bond-order potential (BOP) presented by Murdick \textit{et al.} in 2006\cite{murdick2006analytic} to study the molecular beam epitaxy growth of GaAs MOSFETs. 
Single point calculations for the equation of state, plane decohesion and defect energies have been run with the aid of the Atomic Simulation Environment (ASE) package\cite{ase-paper}.

\subsection{Structural and mechanical properties}

As a sanity check for our MLP, we compute the equation of state and the elastic constants for some stable phases of Ga, As, and GaAs. As a starting point, we use primitive cells obtained from the Materials Project\cite{Jain2013}, and optimize them separately for each potential, to provide a self-consistent reference. 
The results are shown in table \ref{Tab:EOS_Cij}, together with the available experimental values, which the empirical potentials are fitted against. The same set of calculations has been repeated for each potential, and the results of the ANNK and BOP potentials are in agreement with those presented in their respective original papers with the sole exception of the bulk modulus of the ANNK potential for $\alpha$-Ga and Ga-II, which we find to be more than twice as large as the original value reported, a discrepancy whose origin we could not determine.
As expected, our potential is in excellent agreement with the DFT data, while the ANNK and BOP potentials show good agreement with the experimental values of GaAs but are less accurate for single-species Ga and As phases, despite the fact that they were included in the fitting.

\begin{table*}[tbph]
\begin{ruledtabular}
\begin{tabular}{cccccc}
Property            & DFT   & NNP              & ANNK BOP & Murdick BOP & Exp    \\ \hline
\multicolumn{6}{c}{GaAs - ZB}               \\ \hline
V$_0$ [\AA$^3$]     & 23.82 & 23.76 $\pm$ 0.07 & 22.56    & 22.70       & 22.58  \\
E$_0$ [eV/atom]     & -4.04 & -4.04 $\pm$ 0.00 & -3.35    & -3.37       & -3.35  \\
B [GPa]             & 58.82 & 59.75 $\pm$ 0.16 & 71.3     & 73.0        & 74.8   \\
C$_{11}$ [GPa]      & 98.72 & 96.22 $\pm$ 1.25 & 124.96   & 118.89      & 118.1  \\
C$_{12}$ [GPa]      & 41.23 & 46.44 $\pm$ 1.22 & 49.47    & 54.63       & 53.2   \\
C$_{44}$ [GPa]      & 50.92 & 44.09 $\pm$ 0.38 & 39.27    & 47.94       & 59.2   \\ \hline
\multicolumn{6}{c}{GaAs - WZ}               \\ \hline
V$_0$ [\AA$^3$]     & 23.81 & 23.79 $\pm$ 0.06 & 22.56    & 22.70       &   \\
E$_0$ [eV/atom]     & -4.03 & -4.03 $\pm$ 0.00 & -3.35    & -3.37       &   \\
B {[}GPa{]}         & 58.73 & 59.01 $\pm$ 0.73 & 71.25    & 73.00       &   \\ 
\hline\multicolumn{6}{c}{Ga - $\alpha$}     \\ \hline
V$_0$ [\AA$^3$]     & 20.38 & 20.37 $\pm$ 0.02 & 19.27    & 20.88       & 19.58  \\
E$_0$ [eV/atom]     & -2.83 & -2.83 $\pm$ 0.00 & -2.83    & -2.57       & -2.810 \\
B [GPa]             & 46.91 & 47.52 $\pm$ 1.80 & 90.75    & 49.1        & 61.3   \\ \hline
\multicolumn{6}{c}{Ga - II}                 \\ \hline
V$_0$ [\AA$^3$]     & 19.02 & 19.00 $\pm$ 0.07 & 16.53    & 16.71      &        \\
E$_0$ [eV/atom]     & -2.81 & -2.81 $\pm$ 0.00 & -2.86    & -2.60      &        \\
B [GPa]             & 47.57 & 48.25 $\pm$ 1.90 & 350.04   & 98.94      &        \\ \hline
\multicolumn{6}{c}{As}                      \\ \hline
V$_0$ [\AA$^3$]     & 22.42 & 22.42 $\pm$ 0.03 & 19.25    & 19.86      & 21.51  \\
E$_0$ [eV/atom]     & -4.55 & -4.55 $\pm$ 0.00 & -2.91    & -2.94      & -2.9   \\
B [GPa]             & 67.85 & 68.03 $\pm$ 0.37 & 69.03    & 103.20     & 55.6  
\end{tabular}
\end{ruledtabular}
\caption{Comparison of the structural properties between DFT, NNP, ANNK\cite{albe2002modeling} and BOP\cite{murdick2006analytic}. Note that the ANNK and BOP potentials are fitted to reproduce experimental data, while our potential has been fitted on the DFT predictions.\label{Tab:EOS_Cij}}
\end{table*}

\subsection{Defects}

Structure and stability of defects are very important quantities for III/V semiconductors, because of the impact have on electronic properties and device performance. 
In this section we demonstrate the accuracy of MLP prediction for point and planar defects for the stable phases of As, Ga, and GaAs, while also showing the results obtained with the ANNK and BOP potentials. It should be noted that various works in the literature report that the most stable configuration of some of the defects that we present is charged\cite{chiang1975properties,liu1995native,schultz2009simple,malouin2007gallium}. However, we study them in their neutral state, because both the MLP and the empirical potentials do not have any information about the overall charge of the system, but rely only on the nuclear coordinates for their prediction. 
While we could, in principle, train the potential with charged defects instead of the neutral ones, this would be inconsistent with the rest of the bulk structures, that are neutral. This also limits the types of defects that we can study (e.g. surface reconstructions, that often involve macroscopic charge transfer).

\begin{figure}[tbph]
  \centering
  \includegraphics[width=0.95\columnwidth]{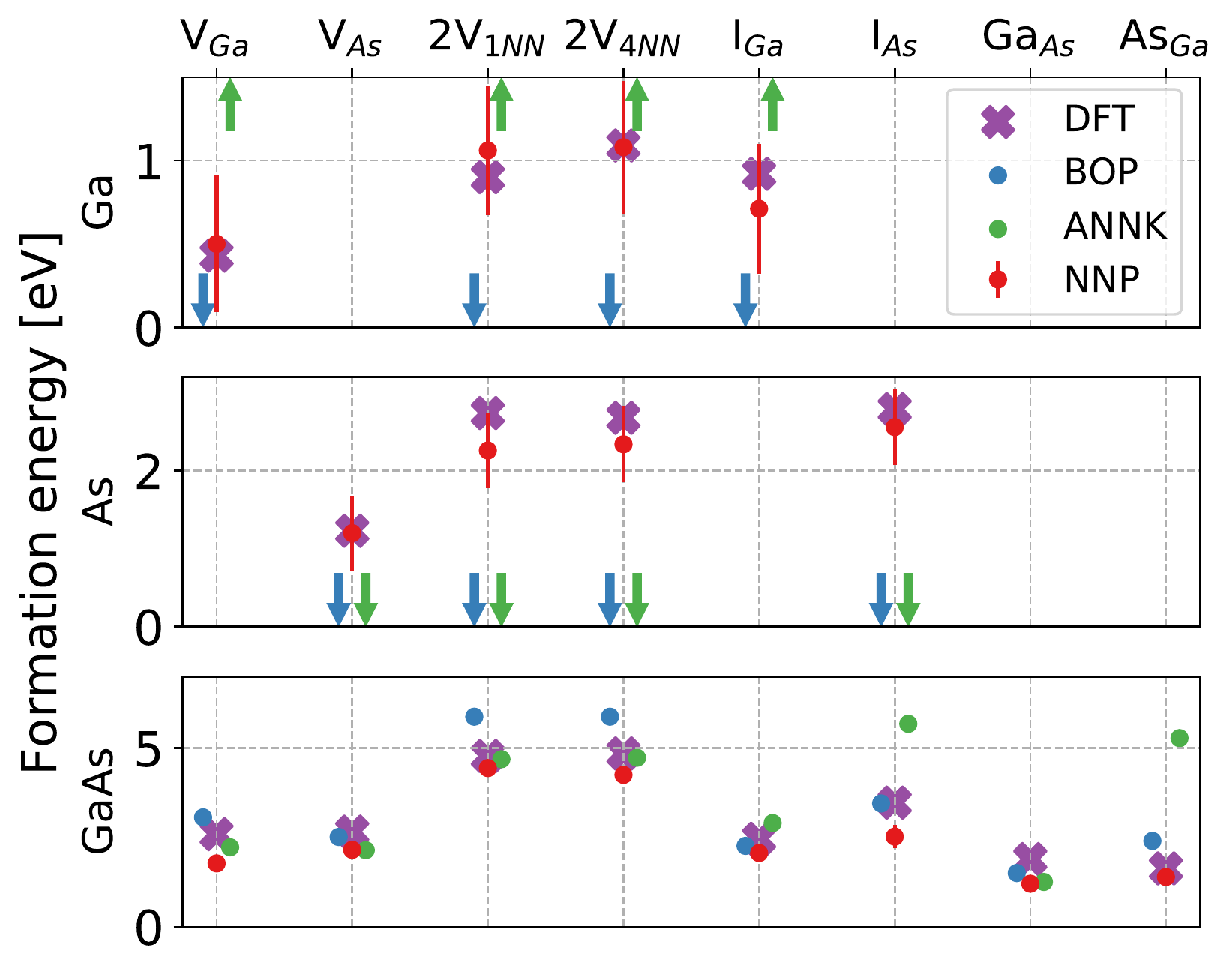}
  \caption{Formation energy of selected defects in bulk Ga, As, GaAs. In the legend, V indicates a vacancy, 2V a divacancy, I an interstitial, and Ga$_{\text{As}}$ is an antisite, where a Ga atom substitutes an As one (and viceversa for As$_{\text{Ga}}$). When multiple defects of the same kind were available, only the lowest-energy one has been presented. The arrows indicate predictions that are far outside the range of reasonable values for the given defects. Numerical values are reported in the \SI, Table S2.}
  \label{fig:all_defects}
\end{figure}

\subsubsection{Point defects}

We compute the formation energies of vacancies, di-vacancies and interstitial atoms for Ga, As and GaAs. For the latter, we also include antisite configurations (substitution of an atom with the other chemical species, e.g. Ga instead of As). 
For each potential we generate the defective supercell at the corresponding equilibrium density, followed by relaxation of the internal coordinates using the BFGS optimization algorithm. 
Therefore, when comparing the various ``relaxed'' configurations, we are effectively observing different minimum energy configurations, each obtained with the corresponding potential.

Since we could not find reference values for the geometry of interstitial atoms in crystalline Ga and As, we generate several possible configurations, and report here only the one that yields the lowest energy of formation with DFT, although we include in the training set all of those that have been created. Similarly, we compute all interstitial configurations that have been reported in GaAs\cite{zollo2003small,schultz2009simple}, but only discuss here what we find to be the most stable structure. Results for the other geometries can be found in the \SI, Table S2.

Results for all defects are summarized in Fig.\ref{fig:all_defects}. The ANNK and BOP potentials both fail to produce meaningful results for defects in As and Ga, yielding extremely high, or negative formation energies -- demonstrating the unphysical results that can be produced by an empirical forcefield outside of the range of configurations it has been fitted for. On the other hand, the predictions for GaAs are closer to the DFT values.
Our MLP can predict with a low error all the formation energies, although it tends to underestimates some particular defects. We also observe that, occasionally, the MLP geometry optimization converges to a structure having a small but significant distortion relative to the DFT geometry, which is associated with a further decrease in energy.  When using the DFT-minimized structures for the comparison, the NNP is able to produce results closer to the DFT references, as shown in the \SI.
Given however that the overall error in terms of energy per atom is much smaller than the overall RMSE of the potential, we found that even adding more reference configurations could not improve the accuracy of the MLP, which underscores the need of including more specific training targets if one wants to achieve the ultimate accuracy in properties that depend on energy differences. 

\subsubsection{Surface energies and reconstructions}

\begin{figure}[tbph]
  \centering
  \includegraphics[width=0.95\columnwidth]{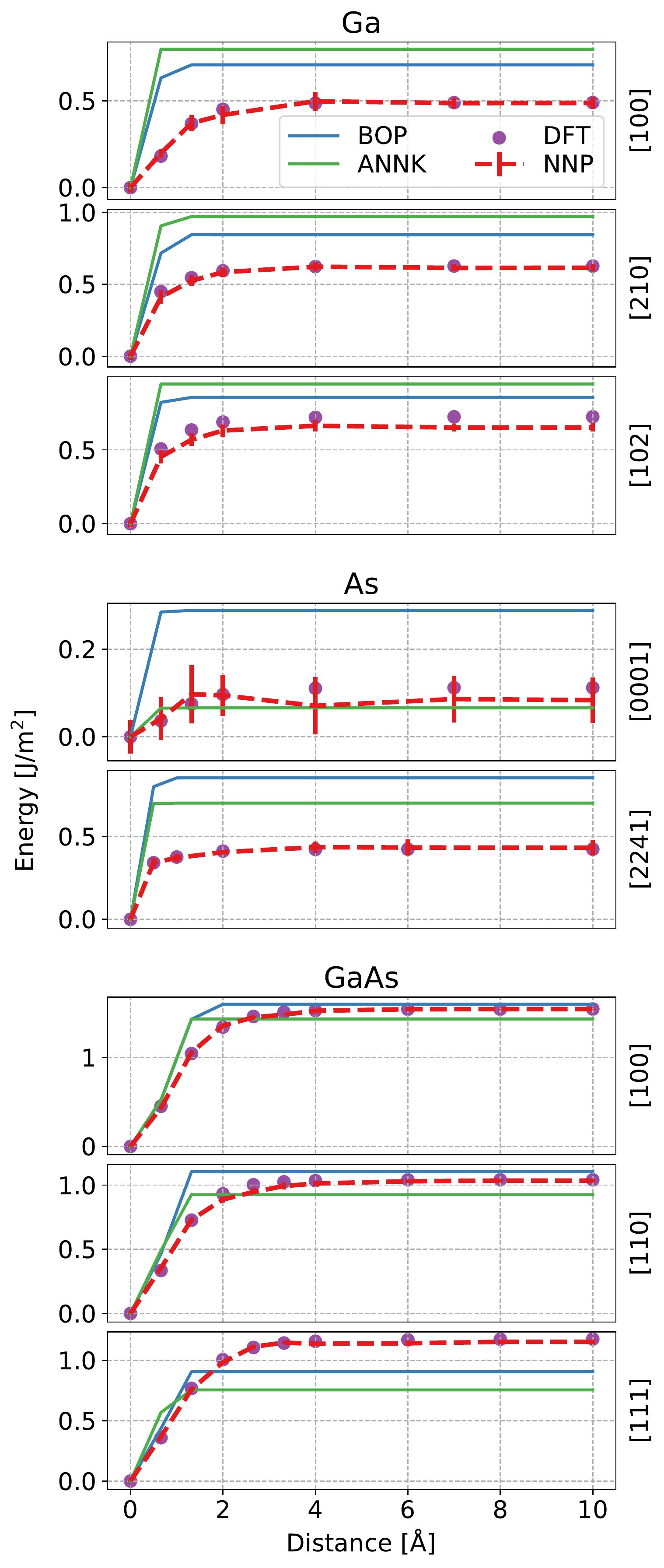}
  \caption{Decohesion energy curves for the main high-symmetry surfaces in Ga, As, and GaAs. Results with BOP and ANNK potentials, our MLP and the DFT reference values are reported. Errorbars from the NNP reflect the distribution of estimates from the calibrated committee model.}
  \label{fig:plane_decohesion}
\end{figure}

Figure \ref{fig:plane_decohesion} reports the rigid decohesion energies for all the stable surfaces of Ga, As, and GaAs, that are relevant to modelling fracture, and the surface-related phenomena that are relevant to modelling the synthesis of III/V nanostructures. Each supercell is computed at the equilibrium density of the corresponding potential. Even though in this case surface energies have reasonable values for all potentials, only the NNP reproduces quantitatively the DFT reference, and avoids an unphysical, near-discontinuous behavior of the decohesion curve. 

However, cleaved surfaces are usually not the most stable structure. The surfaces of semiconductors often undergo complex reconstructions, i.e. the atoms on the surface rearrange themselves and/or bind to one or more adatoms in the presence of a Ga or As atmosphere\cite{ohtake2008surface,moll1996gaas}. Just as for silicon~\cite{gibson1985direct,kitamura1991observation,harrison1976surface}, surface reconstructions in GaAs have been subject of intense experimental and theoretical investigation, and many structures have been proposed and found for each of the high-symmetry orientations, i.e. [100], [110], and polar [111].~\cite{ohtake2008surface,moll1996gaas,chadi1987atomic,biegelsen1990surface,pashley1988structure}

When computing the surface free energy for the reconstructions, we have to account for the variation in stoichiometry of the configuration. We also assume that the surface is allowed to exchange atoms with a reservoir with a given chemical potential. The equilibrium free energy is obtained as

\begin{equation}
    \gamma_{\text{surf}}A = E_{\text{surf}} - \sum_i\mu_iN_i
\end{equation}

where N$_i$ is the number of atoms of the species \textit{i} in the system, and $\mu_i$ its chemical potential in the reservoir. The upper limit of the chemical potentials for each species is that of the respective condensed phase, as $\mu_i < \mu_{i(\textrm{bulk})}$. Since we know that in thermodynamic equilibrium the sum of chemical potentials of As and Ga must be equal to the bulk energy per GaAs pair

\begin{equation}
    \mu_{\textrm{Ga}} + \mu_{\textrm{As}} = \mu_{\textrm{GaAs}} = \mu_{\textrm{Ga(bulk)}} + \mu_{\textrm{As(bulk)}} - \Delta H_f
\end{equation}

we can rewrite our range of chemical potentials in terms of variation of the chemical potential for a single species, which we choose to be As following\cite{moll1996gaas,murdick2006analytic}

\begin{equation}
-\Delta H_f < \mu_{\textrm{As}} - \mu_{\textrm{As(bulk)}} < 0.
\label{eq:mu-as-range}
\end{equation}

Finally, we compute the surface free energy as

\begin{equation}
    \gamma_{\text{surf}}A = E_{\text{surf}} - \mu_{\textrm{GaAs}}N_{\textrm{Ga}} - \mu_{\textrm{As}}(N_{\textrm{As}}-N_{\textrm{Ga}})
\end{equation}

In the case of a cleaved surface we have $(N_{\textrm{As}}-N_{\textrm{Ga}})=0$, thus leaving with a quantity that is independent of the chemical potential.

\begin{figure}[tbph]
  \centering
  \includegraphics[width=0.95\columnwidth]{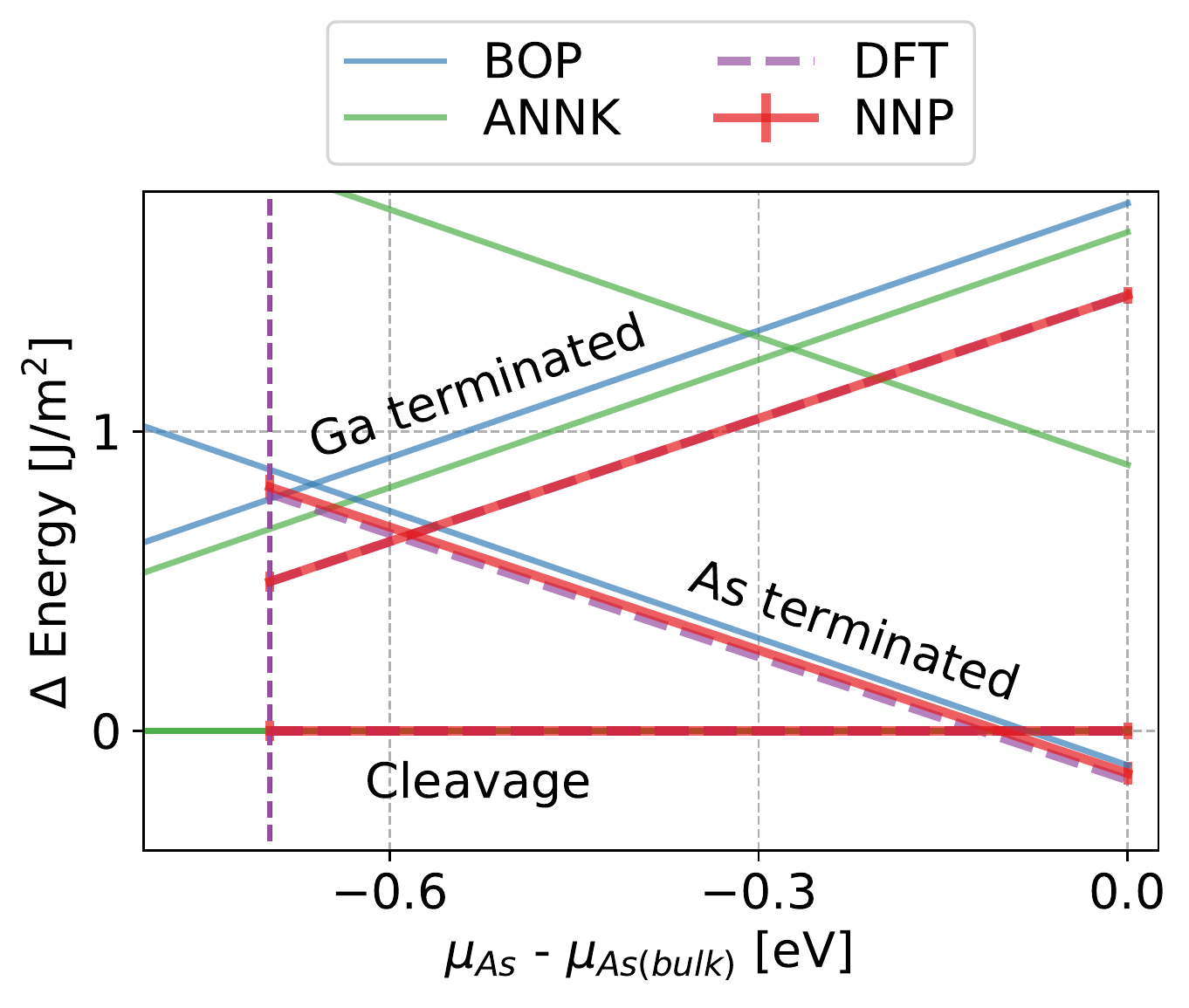}
  \caption{Difference between the surface energy of the various GaAs [110] reconstructions and the cleaved surface plotted against the chemical potential of As. As indicated in Eq.~\eqref{eq:mu-as-range}, the physical values of $\mu_{\ce{As}}$ vary within a range that depends on the potential, from zero down to $-\Delta H_f$, which is $0.7$eV for the NNP and DFT, and approximately $0.9$ for the two empirical potentials. Both the NNP and the BOP potentials recognize the correct stable structures in the observed range, while the ANNK potential finds the cleaved surface as the most stable across all values of the chemical potential. }
  \label{fig:110reconstruction}
\end{figure}

Most of these reconstructions involve charge redistribution between the surface and the bulk. In a typical slab supercell geometry, this requires introducing additional atoms to artificial balance the total charge (e.g. saturating dangling bonds with H atoms), and/or performing DFT simulations for charged systems. 
This poses a challenge for interatomic potentials, such as empirical forcefields and MLPs, whose parametrization relies only on the nuclear coordinates, and do not allow varying the overall charge. While it would be possible to compute MLP results for the [100] and polar [111] surfaces, and compare them with neutral-slab DFT simulations, the results would not be physically significant. As such, we compute and present results for the reconstruction of the [110] surface, the only one which is neutral in every case. 
As shown in Fig.~\ref{fig:110reconstruction}, the MLP reproduces accurately the DFT results; the BOP also predicts qualitatively the correct ordering of surface reconstructions, while the ANNK potential incurs a large error in predicting the stability of the As-terminated reconstruction, and therefore incorrectly predicts the cleaved surface to be the most stable across all values of $\mu_{\ce{As}}$ we consider.

\subsubsection{Generalized stacking fault energy}

\begin{figure}[tbph]
  \centering
  \includegraphics[width=0.95\columnwidth]{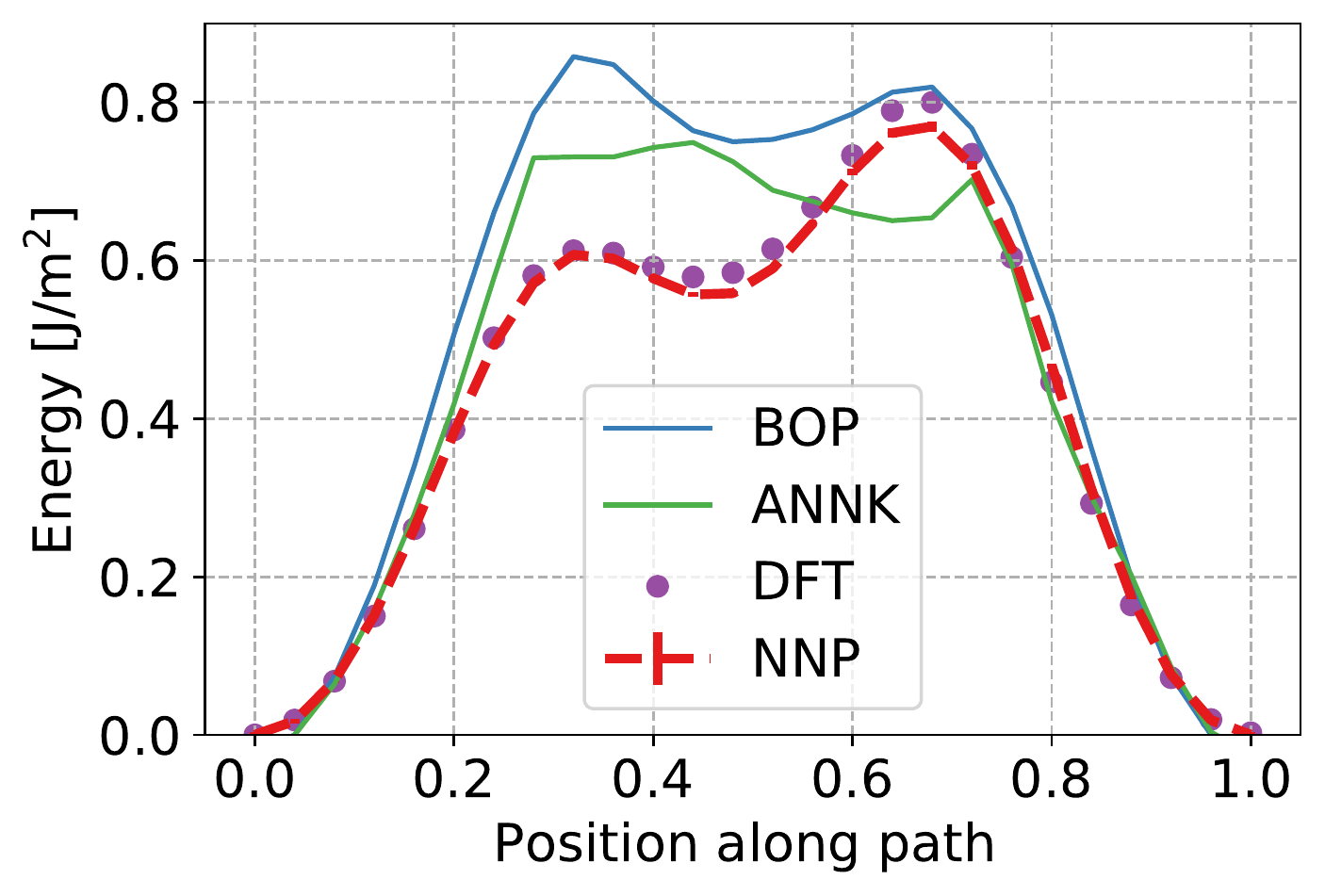}
  \caption{Generalized stacking fault energy profile for the [111] surface gliding along the $\langle$11$\bar{2}\rangle$ direction. The BOP, ANNK and MLP are compared to DFT reference calculations.}
  \label{fig:gsf}
\end{figure}

Surface energies play an important role in the brittle fracture behavior of a material. The generalized stacking fault (GSF) surface, instead, describes the energy cost associated with the sliding of two atomic planes, which is connected to plastic deformation, and the formation and dynamics of dislocations. 
We consider the [111] GSF surface gliding in the $\langle$11$\bar{2}\rangle$ direction. We use a 24 atoms surface and the tilted supercell approach\cite{yin2017comprehensive} to estimate the GSF energy profile (Figure~\ref{fig:gsf}). All the curves exhibit a similar overall glide barrier, but only the MLP reproduces qualitatively and quantitatively the DFT results. The ANNK potential predicts a flat-top, non-smooth GSF profile, while the BOP incorrectly predicts the asymmetry of the path. 

\begin{figure*}[tbph]
  \centering
  \includegraphics[width=1.8\columnwidth]{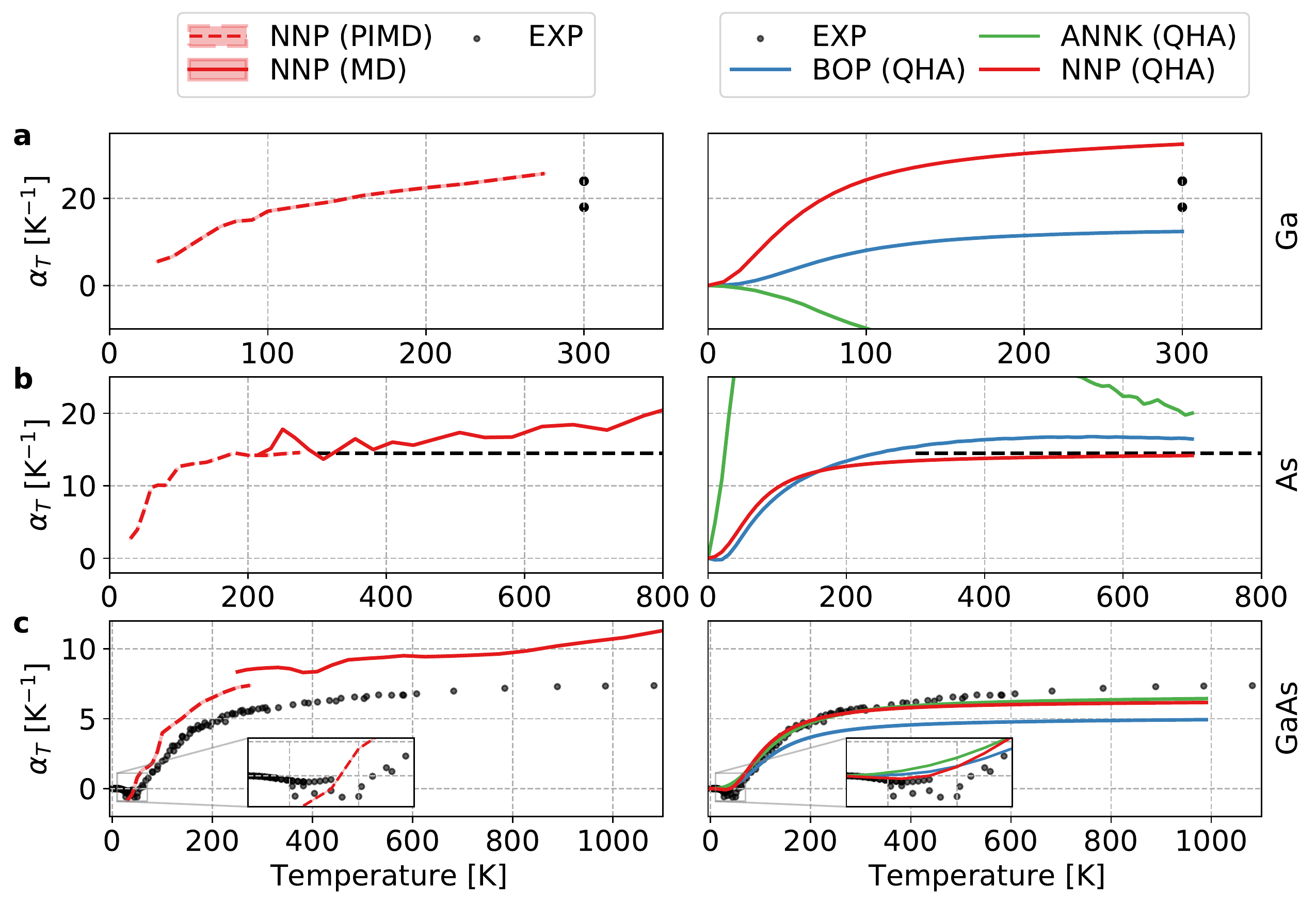}
  \caption{The isotropic thermal expansion coefficient computed for Ga (a), As (b), and GaAs (c). On the left side we provide the results obtained with PIMD (up to 300 K) and MD (from 200 K onward) for the NNP, while on the right side we compare the three potentials at the QHA level. The inset in the (c) panel provides a clear view of the behaviour at low temperature of the three potentials. In this regime, our NNP is the only potential able to recover the negative thermal expansion coefficient. The experimental value of Ga is presented as the range between the maximum observed value and the minimum\cite{galliumazom}, while for As it has been provided as an average over a large range of temperatures As\cite{ananichev2002thermal}. Various sources have been used for the experimental thermal expansion of GaAs\cite{feder1968precision,glazov2000thermal,soma1981phonon,novikova1961investigation,smith1975low,sparks1967thermal} }
  \label{fig:alphaT_all}
\end{figure*}

\section{\label{sec:results}Finite-temperature properties}

Having demonstrated the accuracy of the MLP for quantities that can be computed from static lattice calculations, and for which a direct comparison with DFT reference values is simple, we now move to consider finite-temperature properties,  that require large simulation boxes and long sampling time, and that would be prohibitively demanding when performed with \textit{ab initio} molecular dynamics. 
We investigate a broad temperature range, from 20 to 1600K, that covers both a cryogenic regime, which is well below the Debye temperature and requires a quantum mechanical treatment of the ionic degrees of freedom, up to the melting point of the highest-$T_m$ phase, i.e. GaAs. 
Even though some of the quantities we compute can be obtained with approximate, perturbative methods at smaller computational cost, we report the fully anharmonic estimate using MD and path integral MD simulations, which are made feasible by the use of a MLP.
Even though our results reflect accurately the thermodynamics of the MLP, which in light of the validation in Section \ref{sec:validation} is likely to reproduce the DFT predictions, we expect significant deviations from the experimental values, due to the shortcomings of the reference electronic structure methods. Still, the combination of a MLP and accurate finite-temperature sampling makes it possible to improve substantially the accuracy relative to existing empirical force fields. 
Unless otherwise stated, the uncertainties presented for the properties arise from the finite time of the simulations and have been computed by block averaging the simulations to account for the time correlation of the data.

\begin{figure*}[tbph]
  \centering
  \includegraphics[width=1.8\columnwidth]{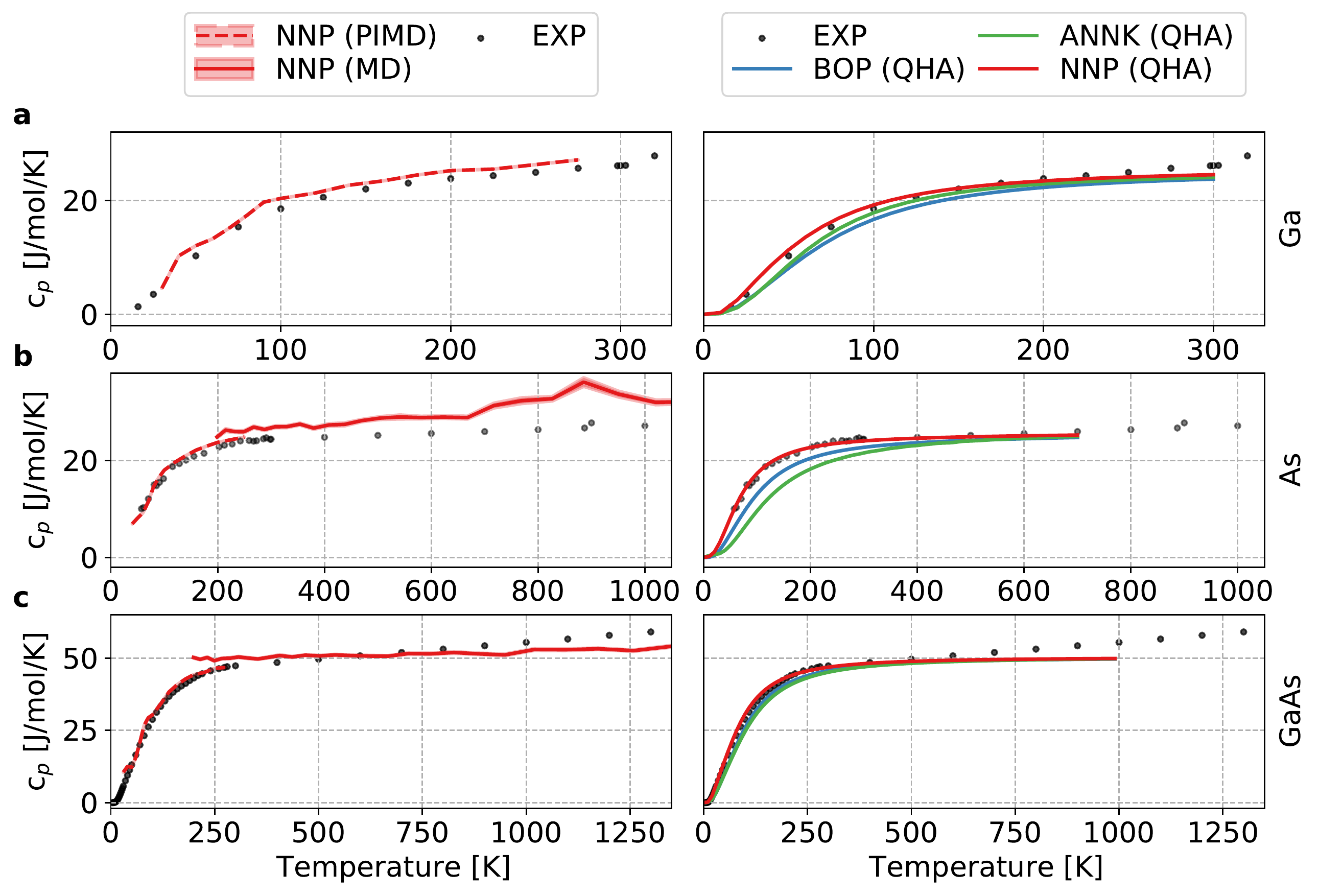}
  \caption{The constant pressure heat capacity coefficient computed for Ga (a), As (b) and GaAs (c) with PIMD (up to 300 K) and MD (from 150 K onward) for the NNP and with the QHA for the BOP and ANNK potentials. MD simulations can only predict the classical value of the heat capacity, whereas with the explicit inclusion of NQEs we can recover the quantum behaviour. Experimental values have been reported for Ga\cite{adams1952heat}, As\cite{anderson1930heat,gokcen1989arsenic}, and GaAs\cite{blakemore1982semiconducting,piesbergen1963durchschnittlichen,lichter1969thermal,cetas1968specific}}
  \label{fig:cp_all}
\end{figure*}

\subsection{Solid properties}

We present the results of simulations of the solid phases for temperatures from 20 K up to the melting point for Ga, As, and GaAs, computed over a fine grid of temperature values. Based on this set of simulations, we compute and discuss bulk thermophysical properties such as heat capacity and thermal expansion for every phase which is stable at room temperature conditions.

The isotropic thermal expansion is computed by comparing the equilibrium volumes between simulations ran at subsequent temperatures (using PIMD and MD simulations separately), while we compute the heat capacity using the variation of the enthalpy with respect to the temperature. %
The same quantities have been also computed with the quasi-harmonic approximation (QHA) as implemented in Phonopy\cite{togo2015first}. In the following figures, we will be presenting on the left side the results obtained with MD for our NNP committee, while on the right side the comparison with the empirical potentials at the QHA level. %

The isotropic thermal expansion coefficients vs temperatures are presented for Ga, As, and GaAs in fig. \ref{fig:alphaT_all} for all three potentials. The ANNK potential shows an unusual profile of the thermal expansion of bulk As and bulk Ga (figs \ref{fig:alphaT_all}a and \ref{fig:alphaT_all}b), while the BOP is able to follow more closely the experimental values. For the case of bulk As, the MD simulations run with the two potentials show that the ANNK potential is unstable when running beyond 800 K, while the BOP is unstable at 1400 K and never undergoes a spontaneous solid-liquid transition. Experimentally, a single result has been found for the isotropic thermal coefficient that can be compared to our analysis, and is given as an average value for temperatures between 300 K and the melting point. Finally, the GaAs results are shown in fig. \ref{fig:alphaT_all}c. GaAs in its zincblende form has a negative thermal expansion coefficient at low temperature\cite{biernacki1989negative}, which is predicted by our potential both in the MD simulations and the QHA, but not by the other models. At higher temperatures, our potential seems to be slightly overestimating the expansion of the solid in the MD simulations. The QHA results follow rather closely those obtained with MD simulations at lower temperatures, while slightly deviating at higher ones, when anharmonic contributions become relevant.

The results concerning the heat capacity converge, as expected, to the corresponding classical value. However, the BOP and the ANNK potentials deviate from the experimental values at low temperatures, particularly for Ga and As (Fig. \ref{fig:cp_all}a and Fig. \ref{fig:cp_all}b respectively). Moreover, as expected, classical MD is not able to reproduce the quantum behaviour of the heat capacity, that can be recovered only by using PIMD simulations, as it can be seen clearly in the calculations run with the NNP for all three phases.

\begin{figure}[btp]
  \centering
  \includegraphics[width=0.95\columnwidth]{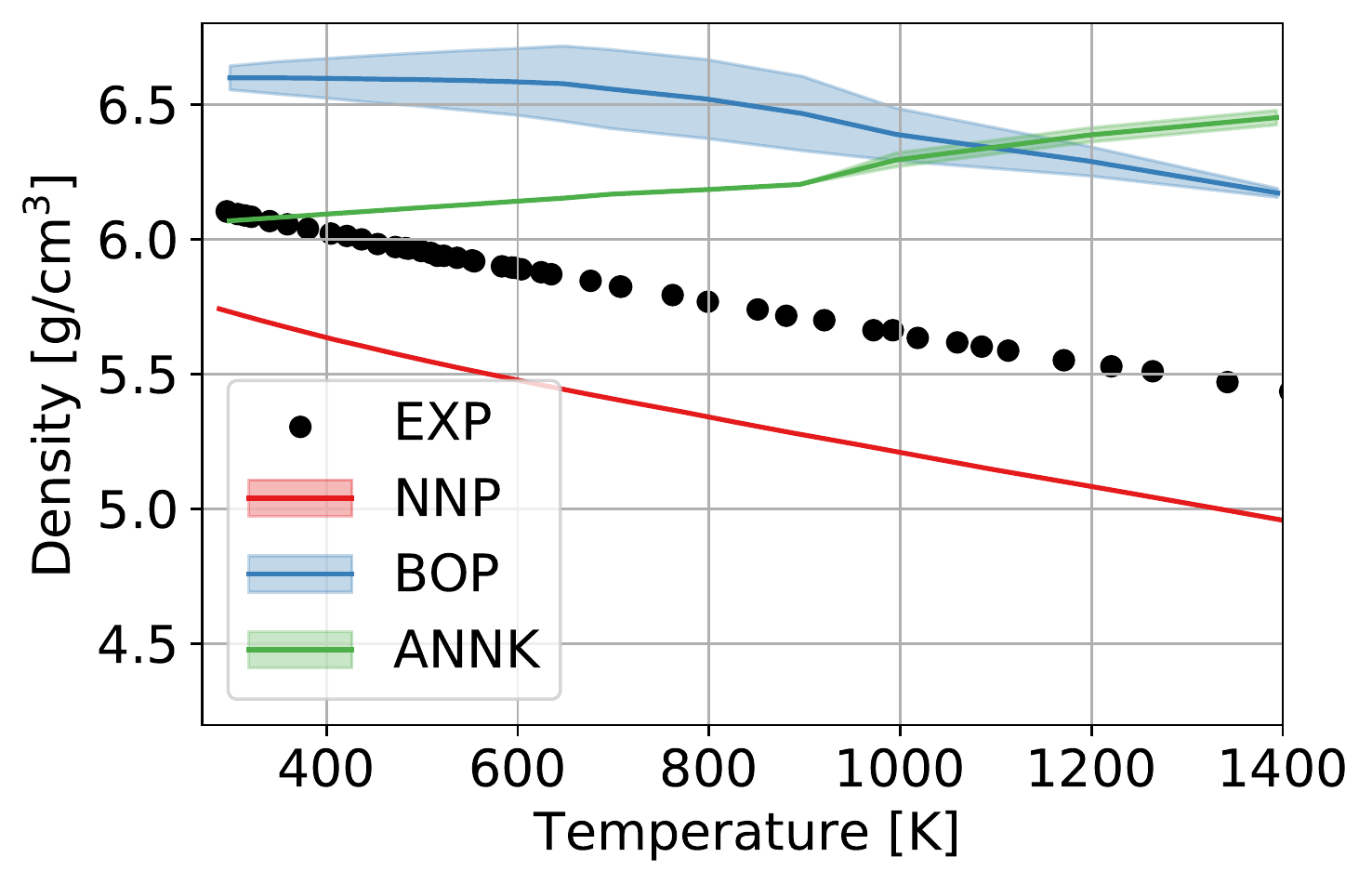}
  \caption{The density of liquid gallium as predicted by the NNP compared to the experimental values. The values from the simulations with the BOP and ANNK potentials are added, but the density refers to the solid phase till 800 K, where a discontinuity is observed for both potentials.}
  \label{fig:density_Ga}
\end{figure}

\begin{figure*}[tbph]
  \centering
  \includegraphics[width=2\columnwidth]{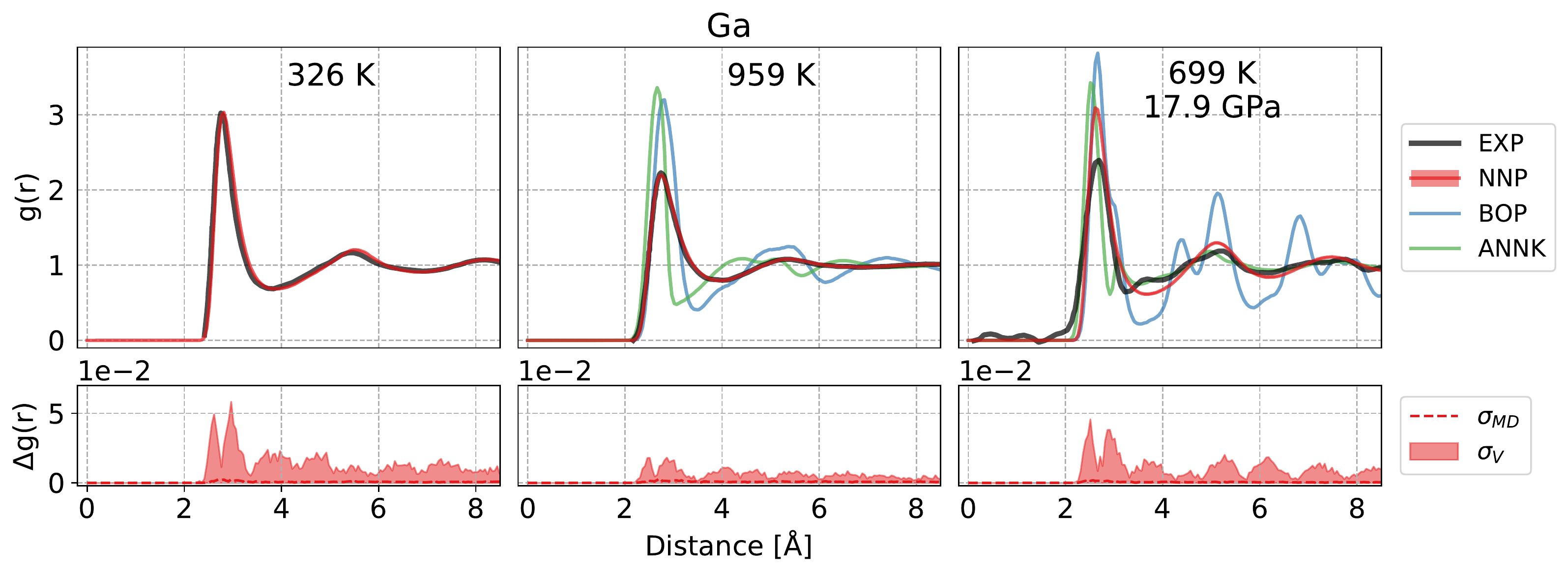}
  \caption{The radial pair distribution function computed for the various potentials and compared to the experimental values at ambient pressure\cite{bellissent1989structure} and at high pressure\cite{drewitt2020structural}. The g(r) of the empirical potentials are not reported in the first panel because the structures remain solid at the reported temperature. The bottom panels present the uncertainty arising from the use of a ML model compared to the statistical uncertainty due to the finite time of the simulation.}
  \label{fig:RDF_Ga}
\end{figure*}

\subsection{Liquid properties}

We turn now our analysis to properties related to the liquid part of the phase diagram, which are investigated using MD simulations of large supercells for long trajectories. In this section we present the results for the density of Ga, the radial pair distribution functions of liquid Ga, As, and GaAs, diffusion coefficients and viscosities of the liquid phases of Ga and GaAs. We also compare the values predicted by our potential with the ones that have been reported experimentally, where available.

The density of liquid Ga is presented in fig \ref{fig:density_Ga}, where it is clear that our potential qualitatively reproduces the experimental values, although underestimating it by about 8\%. \rev{This underestimation may be in part due to the lack of dispersion interactions, that have been found to play an important role in materials composed by row IV elements and above\cite{silv+20jpcb}. 
Investigating the role of dispersion by re-training the NNP against vdW-corrected functionals may be an interesting future line of research.}
Both the BOP and ANNK do not follow even qualitatively the experimental density. Both the empirical potentials are actually solid in the region $\text{T} < 800~\text{K}$ and become liquid only afterwards. A discontinuity in the first derivative of the density can be observed around that temperature for both potentials. As predicted by the thermal expansion calculations of solid Ga (fig. \ref{fig:alphaT_all}a), the ANNK potential actually shows a compression of the box as the temperature increases.

\begin{figure}[tbph]
  \centering
  \includegraphics[width=0.9\columnwidth]{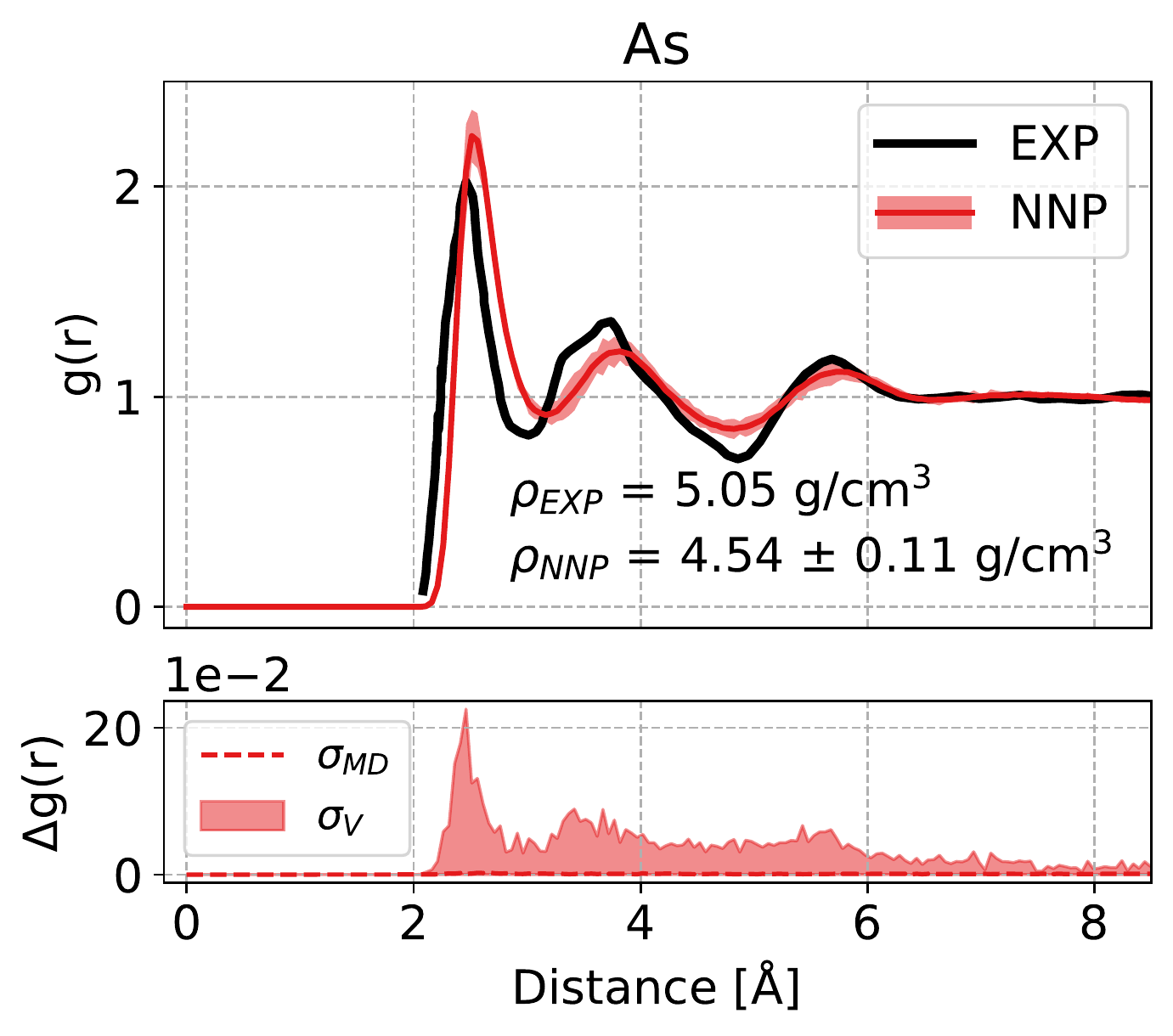}
  \caption{The radial pair distribution function computed for the NNP at 1098 K and 4.8 MPa, compared to the corresponding experimental values available\cite{bellissent1987structure}. The bottom panel presents the uncertainty arising from the use of a ML model compared to the statistical uncertainty due to the finite time of the simulation.}
  \label{fig:RDF_As}
\end{figure}

\begin{figure}[tbph]
  \centering
  \includegraphics[width=0.9\columnwidth]{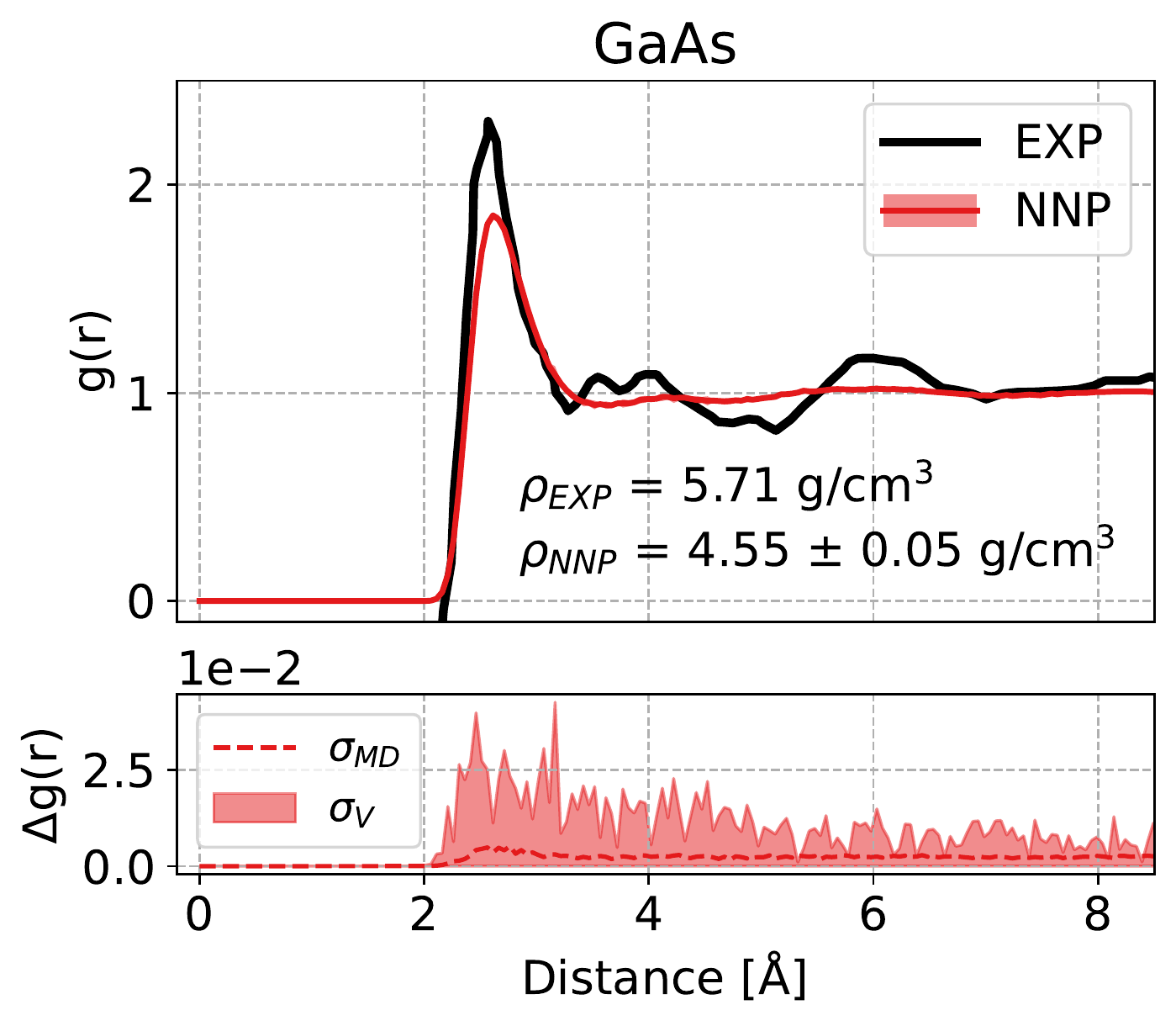}
  \caption{The radial pair distribution function computed for the NNP at 1550 K and compared to the experimental values available\cite{bergman1985atomic}. The empirical potentials have been omitted, since they are solid at this temperature. The bottom panel presents the uncertainty arising from the use of a ML model compared to the statistical uncertainty due to the finite time of the simulation.}
  \label{fig:RDF_GaAs}
\end{figure}

\begin{figure}[tbph]
  \centering
  \includegraphics[width=0.9\columnwidth]{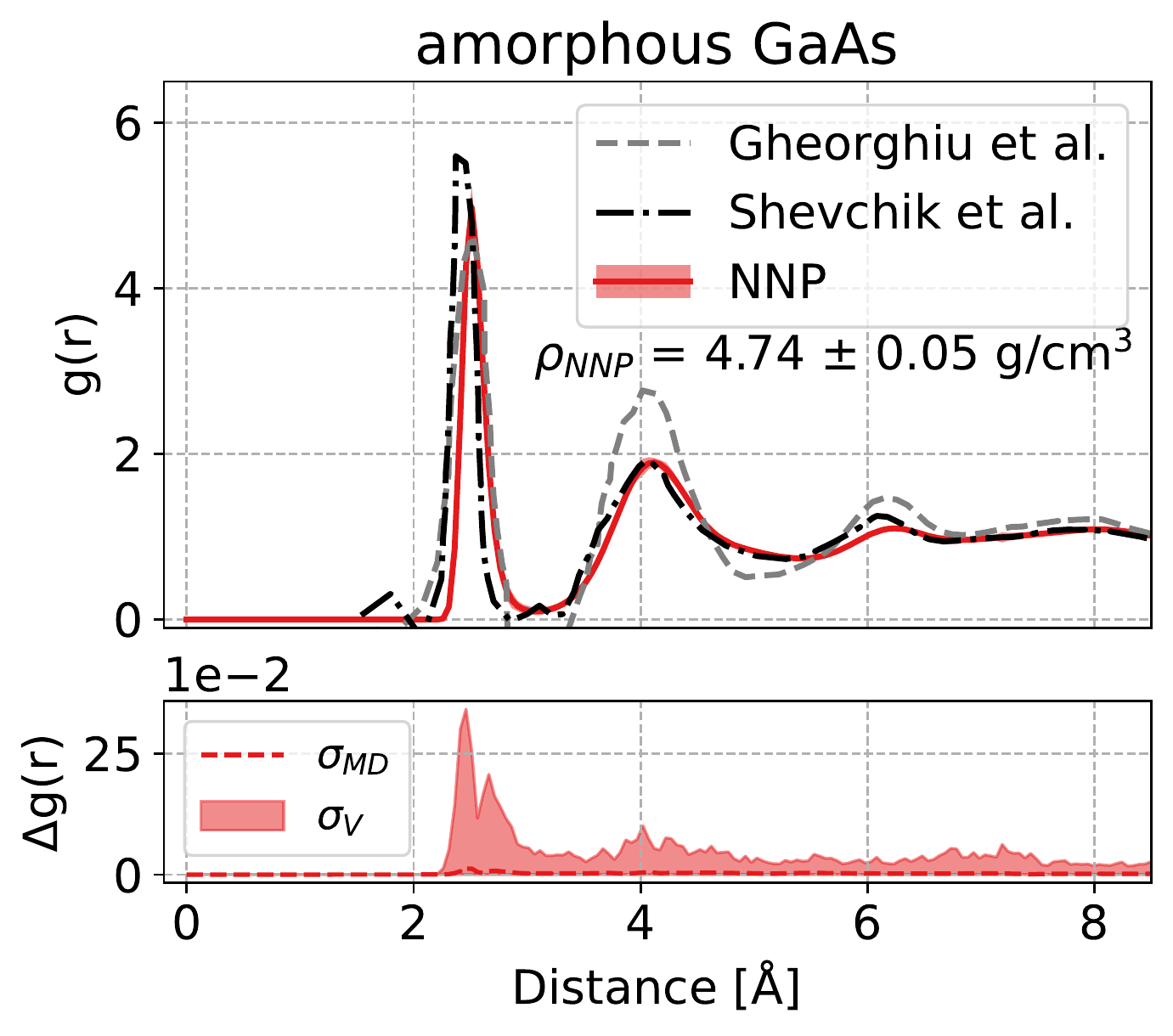}
  \caption{The radial pair distribution function of amorphous GaAs computed for the NNP at 300 K after a slow quenching starting at 1800 K, compared to the experimental values of Gheorghiu \textit{et al.}\cite{gheorghiu1985comparative} and Shevchik \textit{et al.}\cite{shevchik1973structure}. The bottom panel presents the uncertainty arising from the use of a ML model compared to the statistical uncertainty due to the finite time of the simulation.}
  \label{fig:RDF_aGaAs}
\end{figure}

Then, we report the radial pair distribution function, which we will refer to as g(r). We run simulations of liquid Ga at three different conditions in fig. \ref{fig:RDF_Ga}, liquid As in fig. \ref{fig:RDF_As}, liquid GaAs in fig. \ref{fig:RDF_GaAs}\rev{, and amorphous GaAs in fig.~\ref{fig:RDF_aGaAs}} for a comprehensive view of the potential. For As and GaAs, we also provide the equilibrium density at the given temperature. We do not provide the results for the BOP and ANNK potentials in most cases because they are not liquid in the range of temperatures that we have considered for the MD simulations (e.g. the BOP and ANNK melting points of GaAs are reported to be around 1950 K). In these figures we also provide a comparison between the thermodynamic uncertainty obtained by reweighting the trajectories for each potential in the committee (here called $\sigma_V$ following the same notation as Ref.~\citenum{imbalzano2020uncertainty}) and the statistical uncertainty due to the finite time of the simulations (indicated as $\sigma_{MD}$).

In the case of liquid Ga, our potential is able to reproduce with striking accuracy the g(r) both at low and high temperature, similarly to the results of other \textit{ab initio} studies run with GGA\cite{drewitt2020structural,xiong2017temperature} or LDA functionals\cite{niu2020ab}. At higher pressure, we obtain a good agreement with the experimental data, very similar to that of other studies at the GGA kevel\cite{drewitt2020structural}. Both empirical potentials fail to provide a meaningful description of the liquid environment at 959 K, while the ANNK potential has a reasonable, but too ordered, g(r) at high pressure.

Arsenic does not undergo melting at atmospheric pressure, becoming directly gaseous at 887 K. Therefore, in fig. \ref{fig:RDF_As} we run simulations at T = 1098 K and p = 4.8 MPa, where it is liquid, to compare to the g(r) obtained experimentally in the same conditions\cite{bellissent1987structure}. Our prediction is less accurate compared to the Ga one, but we are still able to recover the position of the peaks in the liquid, although the shoulder in the second peak seems to be entirely missing. We are also underestimating the density of the liquid.

The results obtained for liquid GaAs at 1550 K are presented in fig.~\ref{fig:RDF_GaAs}, where we observe a reasonable agreement with the experimental data\cite{bergman1985atomic}, although it is not entirely clear whether the splitting of the peaks in the experiments is a physical feature, possibly due to the undercooling of the sample, or due to the noise. Other \textit{ab initio} simulations in literature also do not show the same splitting of the second peak\cite{zhang1990atomic,molteni1994structure,godlevsky1998ab}. 

\rev{The excessive smoothing of the $g(r)$ of both As and GaAs, and the underestimation of the density, are probably a reflection of the limitations of the ab initio reference rather than of the NNP, as evidenced by the small estimated $\sigma_V$. As in the case of Ga, incorporating dispersion interactions might be a possible strategy to improve the accuracy of DFT energetics.} 

\rev{Finally, we provide predictions of the g(r) for amorphous GaAs, which is not included in the training set. We prepare the cell by first running 5 trajectories with different initial structures made of 1000 atoms where we quench the liquid from 1800 K to 300 K over 1 ns. 
Then, we compute the g(r) on 1 ns-long simulations of the final structure, at 300 K. The results presented in fig. \ref{fig:RDF_aGaAs} refer to the average g(r) of the 5 different simulations, compared to the experimental results of Gheorghiu \textit{et al.}\cite{gheorghiu1985comparative} and Shevchik \textit{et al.}\cite{shevchik1973structure}. Overall, we find a good agreement with the experimental values, with very similar positions of the peaks. We also observe that the uncertainty over the energies is, on average, only twice as large as the same uncertainty computed for liquid GaAs, which translates in an uncertainty in the g(r) that is larger, but still negligible. In fact, the uncertainty of the g(r) of amorphous GaAs is comparable with the one we obtais for liquid As, which is explicitly included in the training set.
Overall, we believe that the potential is able to produce accurate results for the amorphous system, despite the lack of dedicated structures in the training set. 
For a study dedicated to the amorphous phases, however, we would recommend to extend the training set incorporating explicitly amorphous structures.}

\begin{figure}[tbph]
  \centering
  \includegraphics[width=0.95\columnwidth]{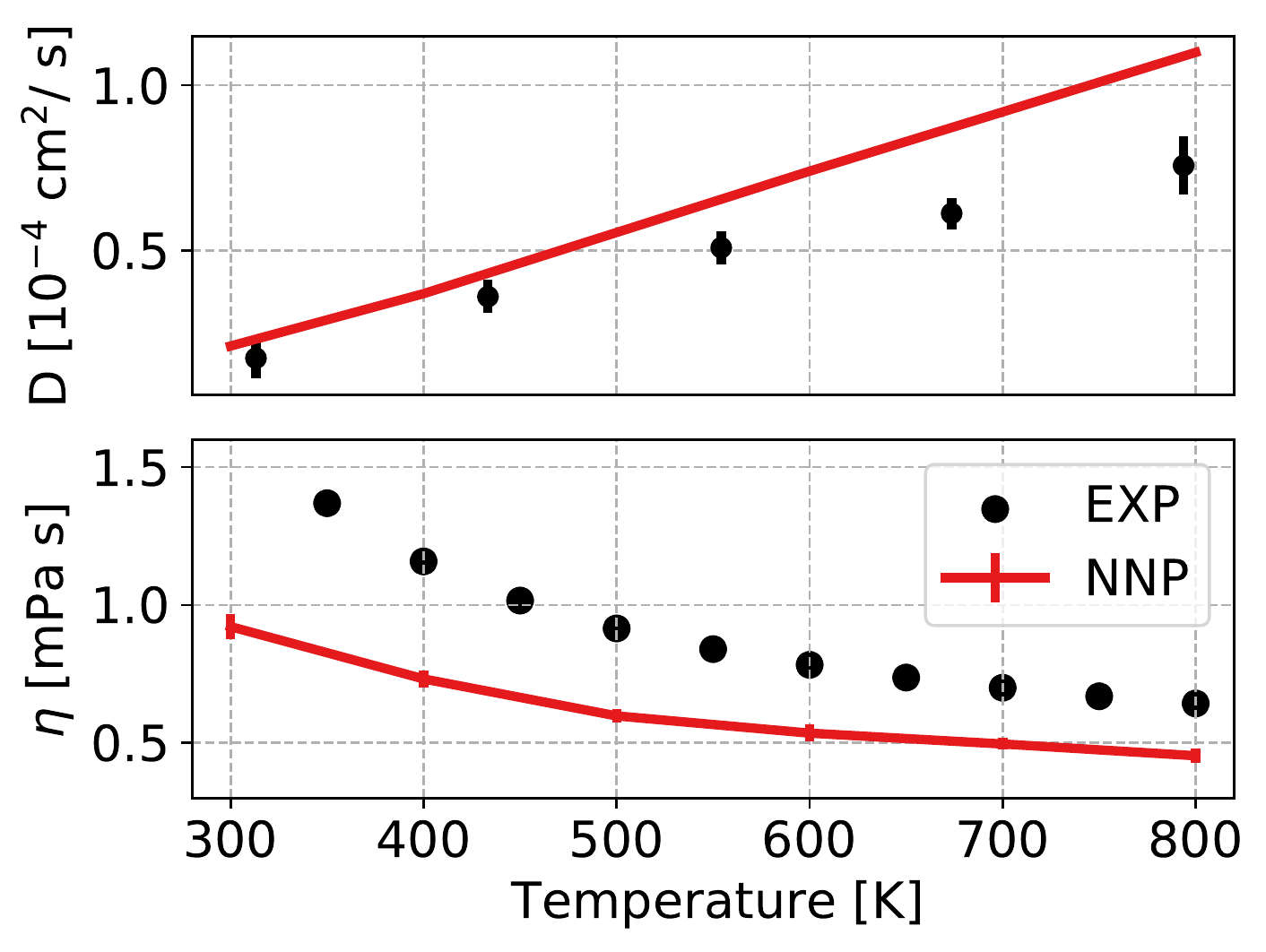}
  \caption{The diffusion (top) and the viscosity (bottom) computed for liquid gallium at increasing temperatures. 20 simulations have been run for each point and the spread in the predicted values is reported with the errorbars. Although the viscosity and the diffusion are related, we have used experimental values reported from separate sources for the viscosity\cite{assael2012reference} and the diffusion\cite{blagoveshchenskii2015self}}
  \label{fig:visc_Ga}
\end{figure}

\begin{figure}[tbph]
  \centering
  \includegraphics[width=0.95\columnwidth]{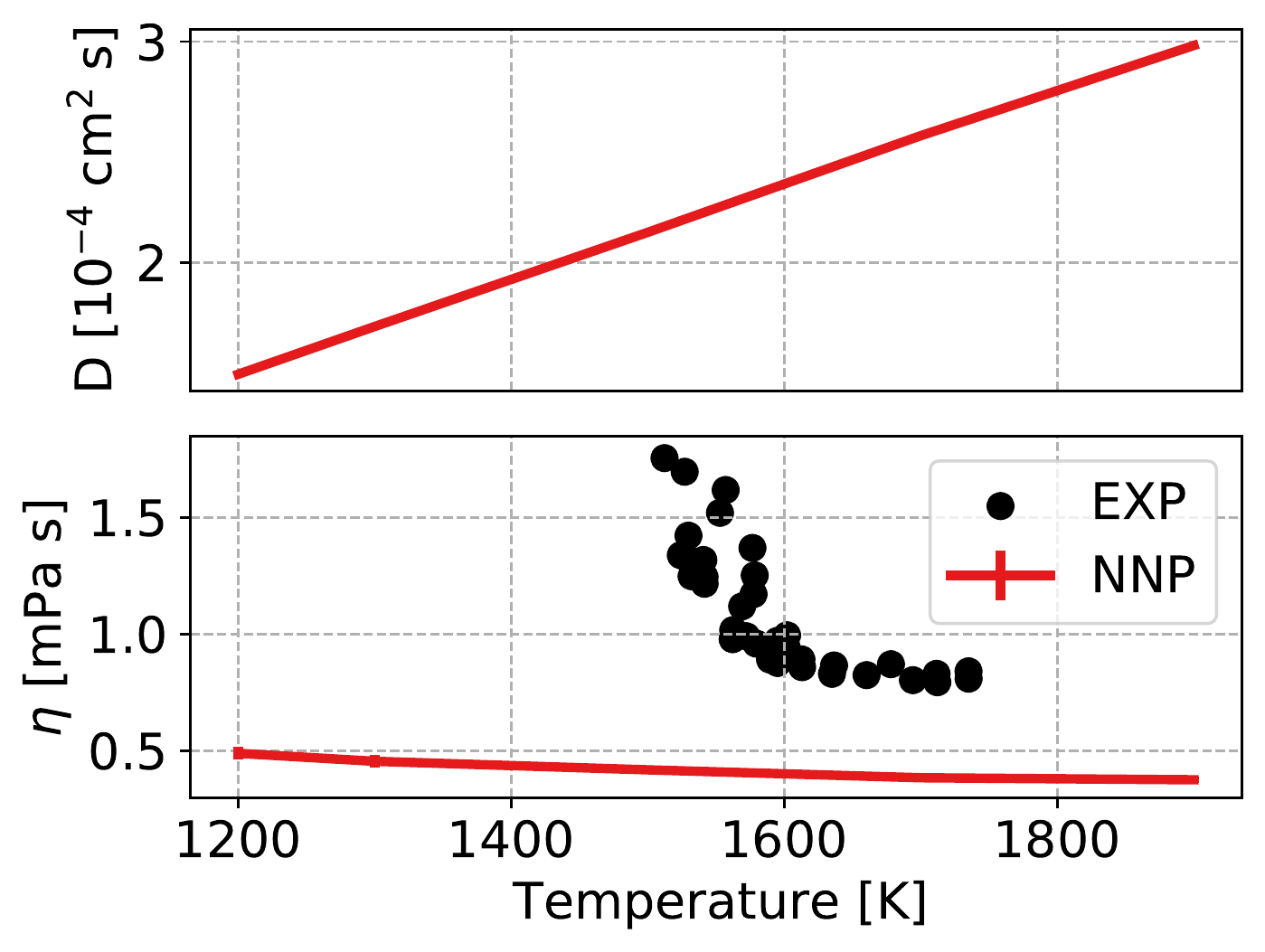}
  \caption{The diffusion (top) and the viscosity (bottom) computed for liquid gallium arsenide at increasing temperatures. 20 simulations have been run for each point and the spread in the predicted values is reported with the errorbars. The simulations are compared to the reported experimental values for the viscosity\cite{kakimoto1987temperature}, whereas no direct measurement of the diffusion has been found in literature.}
  \label{fig:visc_GaAs}
\end{figure}

Having investigated the thermodynamic properties of the bulk liquids, we now consider the diffusion coefficients and the viscosities for the liquid phases of Ga and GaAs. To obtain these, we run several simulations with cubic boxes with a side of 30 \AA, relaxed at the equilibrium density. %
At each temperature we run 20 simulations starting from different initial configurations (at equilibrium density) for 200 ps each in the NVT ensemble using a weak SVR thermostat\cite{bussi2007canonical}. We compute the mean square displacement as an average over the 20 trajectories and obtain the diffusion coefficient for the finite-size system (which we indicate with the PBC subscript) with adequate statistics.
\begin{equation}
    D_{\text{PBC}} = \lim_{t \to \infty} \frac{1}{6t} \langle \sum_{j=1}^{N} (r_{j,i}(t) - r_{j,i}(0))^2\rangle
\end{equation}

\begin{figure}[tbph]
  \centering
  \includegraphics[width=0.95\columnwidth]{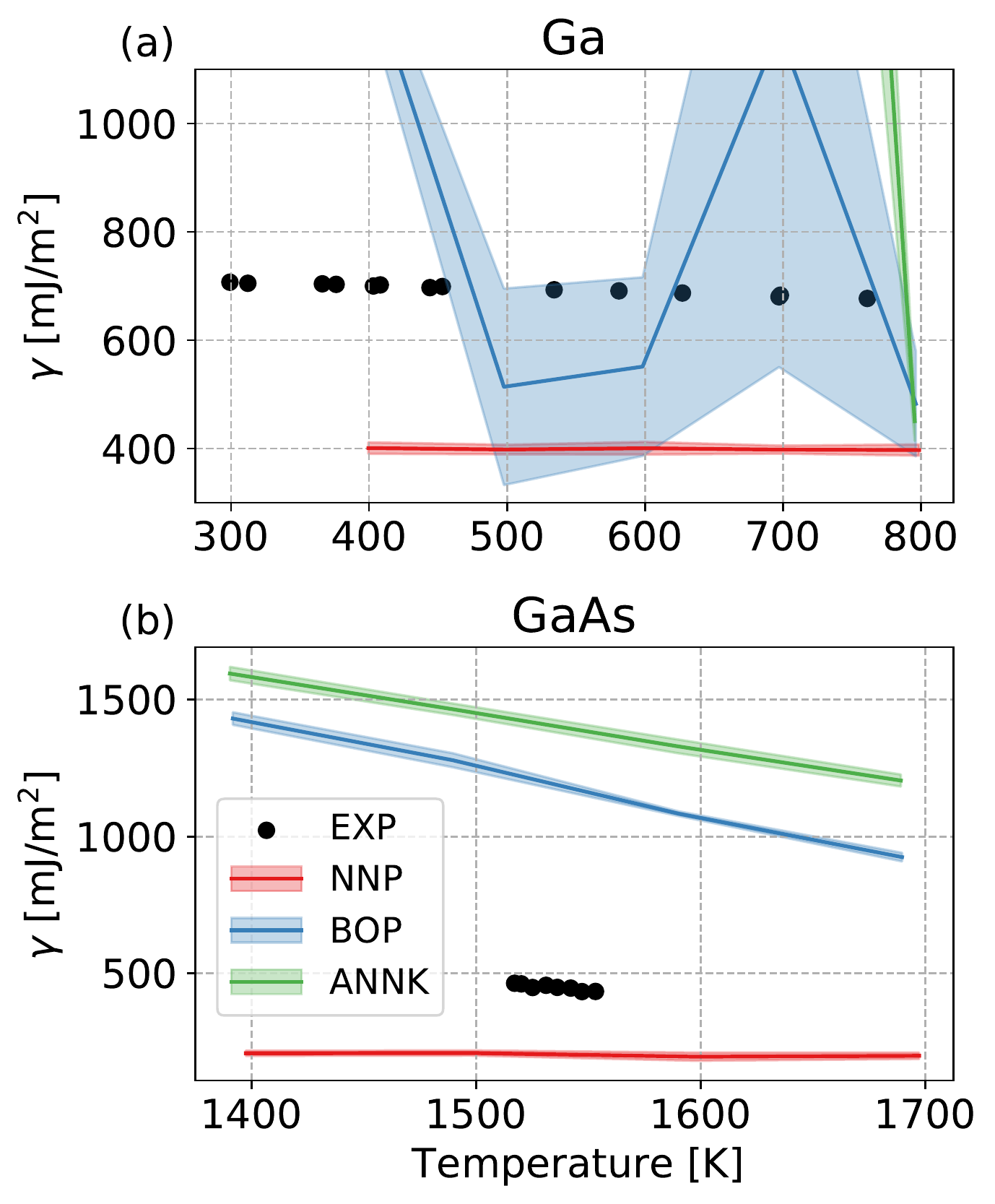}
  \caption{The surface tension of Ga in panel (a) and GaAs in panel (b) for increasing temperature, compared to the experimental values\cite{hardy1985surface,shetty1990surface}. As previously mentioned, the empirical potentials tend to overestimate the melting point, so the results with these potentials refer to the solid phase}
  \label{fig:gamma_all}
\end{figure}

Since the diffusion coefficient is known to be heavily affected by the size of the box\cite{dunweg1993molecular}, we determine the diffusion coefficient of the infinite bulk system by adding the correction factor computed by Yeh and Hummer\cite{yeh2004system}, as
\begin{equation}
    D_\infty = D_{\text{PBC}} + \frac{\xi k_B T}{6\pi \eta L}
\end{equation}
where $\xi$ is a dimensionless constant equal to 2.837297 for cubic simulations boxes, $\eta$ is the viscosity and L is the side of the box. The viscosity, which is almost independent from the box size\cite{yeh2004system,moultos2016system,jamali2018finite}, is obtained from the autocorrelation function of the off-diagonal elements of the stress tensor computed in the same simulation of the diffusion, as 
\begin{equation}
    \eta = \frac{V}{k_BT}\int^{\infty}_{0}\langle P_{\alpha\beta}(t)\cdot P_{\alpha\beta}(0)\rangle \textrm{d}t.
\end{equation}
An alternative method to compute the diffusion coefficient for the infinite bulk system is to compute the coefficient for supercells of increasing size, then extrapolate the value for an infinite supercell\cite{jamali2018finite}. Therefore, we have run additional calculations for smaller cells, to compare the values obtained with the two methods and found them to be in good agreement, as seen in Fig. S1 of the \SI.

While the MLP  consistently underestimates the viscosity (and conversely overestimates the diffusion coefficient), we are able to recover the qualitative behaviour at lower temperature for gallium, as seen in fig. \ref{fig:visc_Ga}. \rev{The underestimation of the viscosity at a given temperature is to be expected given the lower value of the melting point, that we discuss next.
A large underestimation of the viscosity is also observed in the case of GaAs (fig. \ref{fig:visc_GaAs}), which also has a much lower melting point (1200\,K against 1550~K observed experimentally). 
However, the fact that even at the lowest temperature we do not observe the sharp increase in viscosity that is observed in experiments when approaching the melting point suggests that our MLP should be used with care when investigating dynamical properties for molten GaAs.}

Finally, having reported on the bulk properties of the liquids, we compute their surface tension, that we obtain by running 1 ns long simulations of the interface between bulk liquid and vacuum in a large orthorhombic supercell with 1568 atoms for Ga and 1728 atoms for GaAs, with approximate dimensions of 32x32x100 \AA~at varying temperatures. To estimate the surface tension we have used its relation to the diagonal elements of the stress tensor for the described box, as in eq. \ref{eq:surface}, where the 1/2 factor at the beginning accounts for the presence of two interfaces between liquid and vapour.
\begin{equation}\label{eq:surface}
    \gamma = \frac{1}{2} L_z [P_{zz} - \frac{1}{2}(P_{xx}+P_{yy})]
\end{equation}
Our NNP seems to underestimate the surface tension for both Ga and GaAs, as seen in fig. \ref{fig:gamma_all}. To investigate the discrepancy, we check additional structures related to these trajectories and find errors of 1 to 2 meV/atom between our NNP and the DFT results, well below the overall RMSE of the potential, suggesting that the discrepancy might be due to the reference calculations and not to the accuracy of the fit.

\subsection{Binary phase diagram}
In the introduction we mentioned our aim to produce an accurate and transferable potential. Until now we have computed a number of properties with the purpose of showing its accuracy, albeit limited by the underlying DFT reference. Here we want to provide a compelling proof of the transferability of the potential, which is of utter importance when studying technologically relevant phenomena in varying conditions of temperature, pressure, and stoichiometry.

\begin{figure}[tbph]
  \centering
  \includegraphics[width=1.0\linewidth]{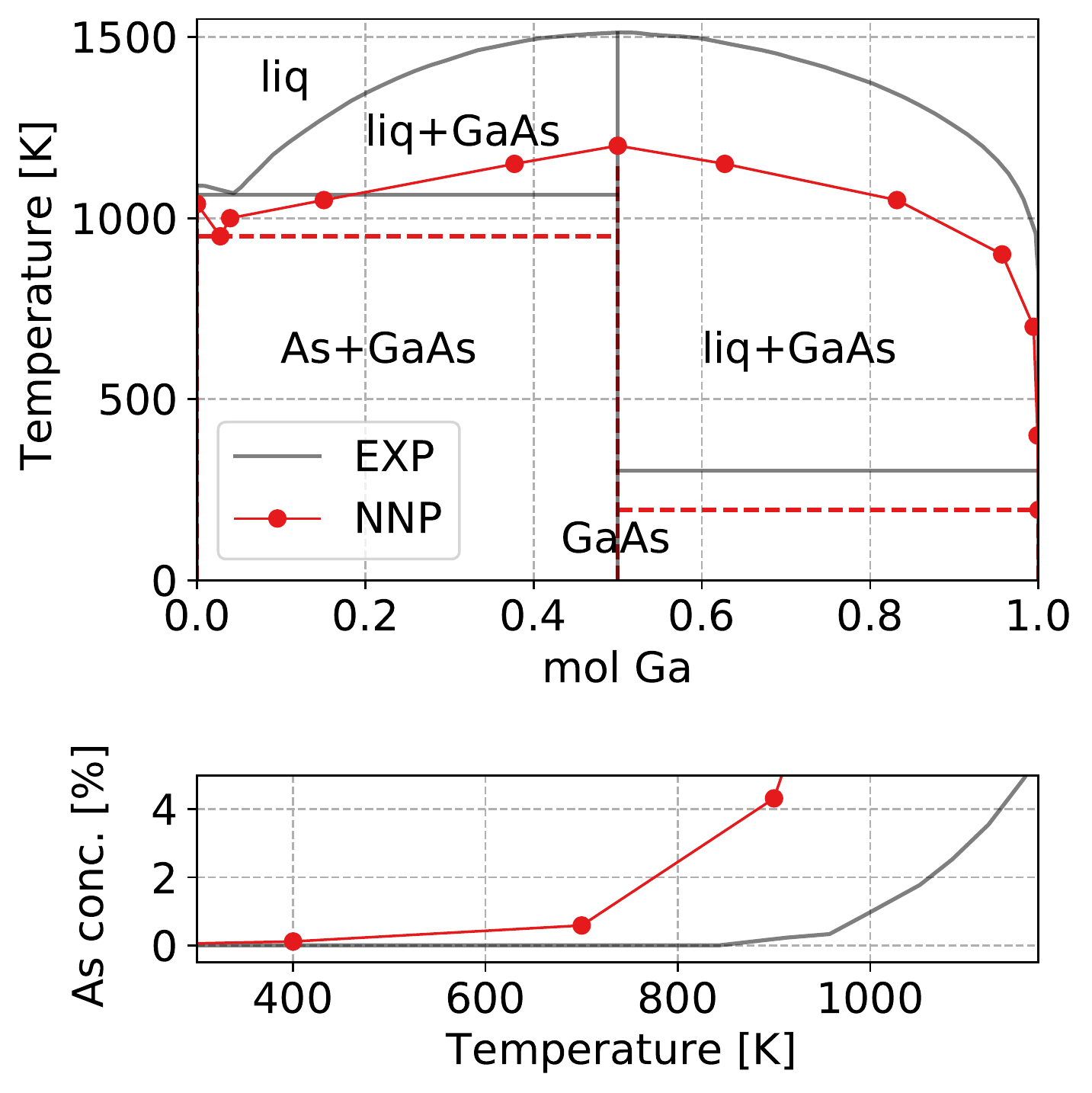}
  \caption{Top: the binary phase diagram for GaAs as predicted from our NNP, compared to the experimental one. We used interface pinning simulations to find the melting point for the pure Ga, As, and GaAs cases, while the other points have been measured using mixed Monte Carlo - MD simulations at different stoichiometries and temperatures. Bottom: saturation concentration of As in liquid Ga as a function of temperature in the region of low temperatures.}
  \label{fig:binary_PD}
\end{figure}

\begin{figure}[tbph]
  \centering
  \includegraphics[width=0.95\linewidth]{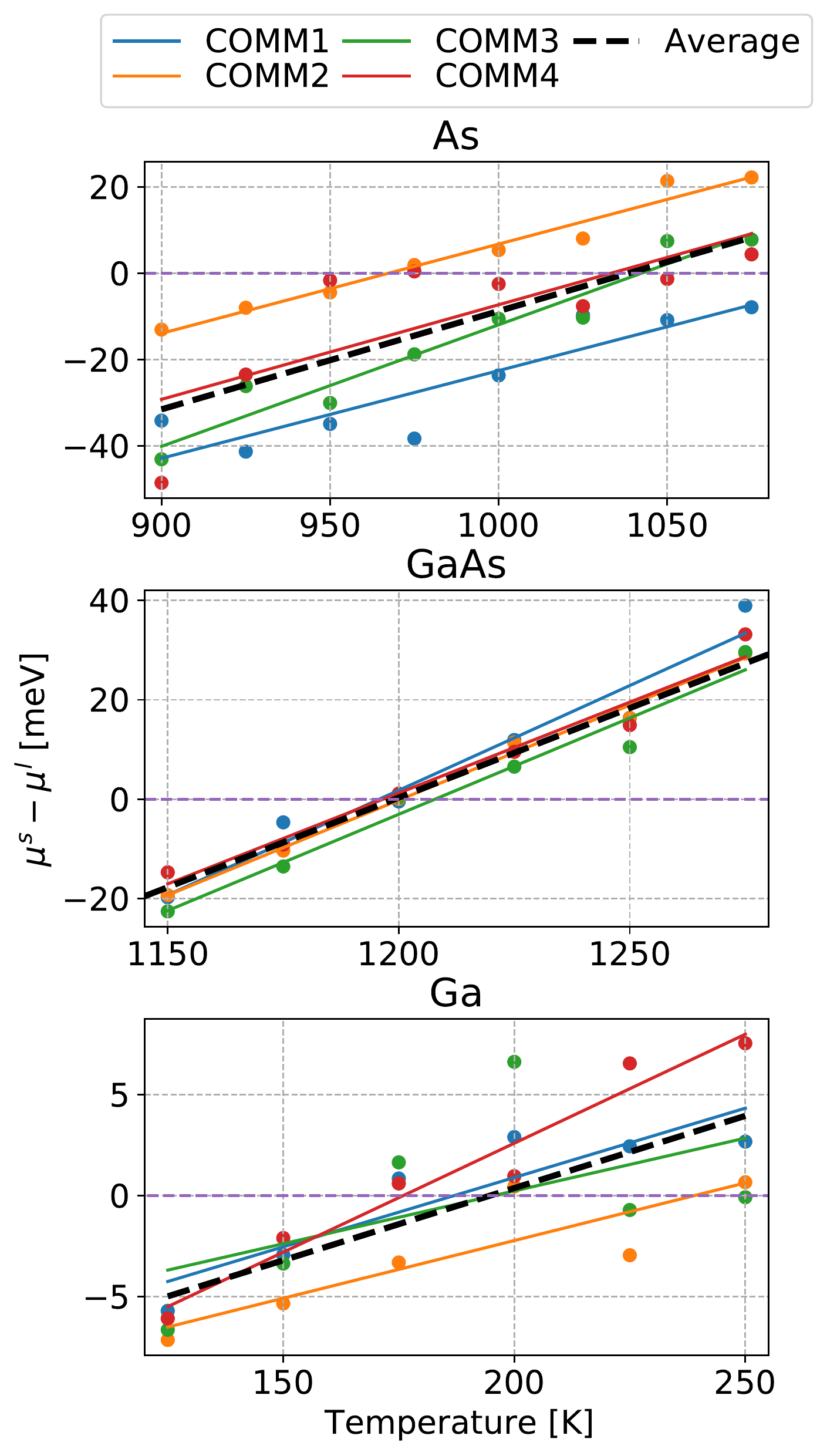}
  \caption{Determination of the error in the melting point of As, GaAs, Ga associated with the NNP fit, using the uncertainty estimation scheme in Ref.~\cite{imbalzano2020uncertainty}. The chemical potentials at each temperature are computed separately for each member of the NNP committee, using a cumulant expansion scheme that makes it possible to obtain the four values by reweighting the trajectory driven by the committee average. This approach makes it possible to estimate the melting point of each potential, and determine the uncertainty in melting point by the spread in the four predictions. }
  \label{fig:MP_uncertainty}
\end{figure}

Providing a full description of the phase diagram is a definitive test of reliability of the potential, since not only we are performing simulations at different stoichiometries, but every simulation that we run contains both solid and liquid bulk, together with their interface. 

In figure \ref{fig:binary_PD} we show our predicted phase diagram, compared to the experimental one. At first glance we observe a good agreement in the shape of the two curves, with a low solubility of As predicted at low temperatures in the high-Ga region (highlighted in the bottom figure), followed by an almost flat central part. We also observe an eutectic point at T = 950 K and $x=0.03$, not far from the experimental value of T = 1083 K and $x=0.05$. The melting points are predicted to be 1039 K, 1200 K, and 195 K for As, GaAs, and Ga respectively, which are in relatively good agreement with the experimental values of 1090 K, 1511 K, and 303 K. 
\rev{It is important to stress that the discrepancy is probably due, in large part, to the underlying electronic-structure reference. In fig~\ref{fig:MP_uncertainty} we demonstrate the use of the thermodynamic uncertainty quantification scheme from Ref.~\citenum{imbalzano2020uncertainty} to determine the error due to the fit of the NNP. We find $1039 \pm  51$ K, $1200 \pm 5K$, $195 \pm 24 K$ for As, GaAs and Ga: except for the case of As, the error associated with the machine-learning approximation is a small fraction of the discrepancy with experiments.  
}

The points shown in figure \ref{fig:binary_PD} have been computed with two different methods. The melting point of pure Ga, As, and GaAs is obtained with the interface pinning method, as described by Pedersen \textit{et al.}\cite{pedersen2013computing}. The remaining points in the liquidus are obtained by running large supercells at various temperatures and stoichiometries, measuring the concentration of the two species in the liquid at equilibrium. In order to speed up the equilibration of the concentrations we add a Monte Carlo step on top of the MD calculation.

For the interface pinning simulations we first determine an optimal collective variable that can distinguish solid and liquid phases, and then run multiple simulations at regular temperature intervals for a large supercell in the Np$_z$T ensemble. To run these trajectories, we use the open source PLUMED library\cite{bonomi2019promoting,tribello2014plumed} to add the bias potential, in addition to i-PI and LAMMPS. After obtaining the mean value of the collective variable at each temperature, we determine the melting point by fitting the chemical potentials to a line. The temperature at which we find a chemical potential of $\mu = 0$ is the melting point of the system\cite{pedersen2013direct}. To compute these trajectories we use the locally averaged Steinhardt parameters introduced by Lechner and Dellago\cite{steinhardt1983bond,lechner2008accurate} q4 (for As and GaAs) and q6 (for Ga) as collective variables for the system.

For intermediate concentrations, we use a mixed Monte Carlo - MD scheme, implemented in i-PI. At every MD step, we attempt to swap 50 (on average) random Ga-As pairs in the system. The particle exchange is then accepted or rejected using a Metropolis criterion. The supercells used in this case are composed by 50\% solid GaAs and 50\% liquid \gaasx. The stoichiometry of the liquid is determined such that the total stoichiometry of the system varies between $0.25 < x < 0.75$. The simulations are divided in a first NpT part, for 10 ps, to find the equilibrium density for the solid, and a second Np$_z$T part, run for 200 ps. In this second trajectory, we allow the system to equilibrate for the first 100 ps, and then measure the average concentration of As and Ga in the liquid for the remaining 100 ps. 

This method works without the need to introduce an external potential to pin the interface because we are considering a binary mixture\cite{baldi2020atomistic}. For an elemental system, the chemical potential between the solid and liquid at the melting point is 0, hence the need to introduce the bias potential to avoid a random walk of the interface, which could result in complete freezing or melting. For the mixture, however, the curvature of the free energy at the interface depends on the composition of the two coexisting phases as
\begin{equation}
    \left(\frac{\delta^2G}{\delta f^2}\right)_{p,T,x} = 
    \frac{(x_s - x_l)^3\mu_l''(x_l)\mu_s''(x_s)}{(x_s - x)\mu_s''(x_s) + (x - x_l)\mu_l''(x_l)},
\end{equation}
where $f$ represents the fraction of solid phase in the system, $x_s$, $x_l$ and $x$ are the compositions of the solid, the liquid, and the overall system, and $\mu_{l,s}$ is the chemical potential of the liquid and the solid, respectively.  Thus, in any case in which solid and liquid have different equilibrium composition, there is a positive curvature that acts as a restoring force against fluctuations of the dividing surface, acting effectively as a pinning potential that keeps the solid fraction fluctuate around the value consistent with the lever rule. Measuring the mean composition of the two phases in equilibrium makes it possible to determine the position of the solidus and the liquidus. The derivation is provided in Ref.~\citenum{baldi2020atomistic}.

\subsection{Beyond potentials}

\begin{figure}[tbph]
  \centering
  \includegraphics[width=\columnwidth]{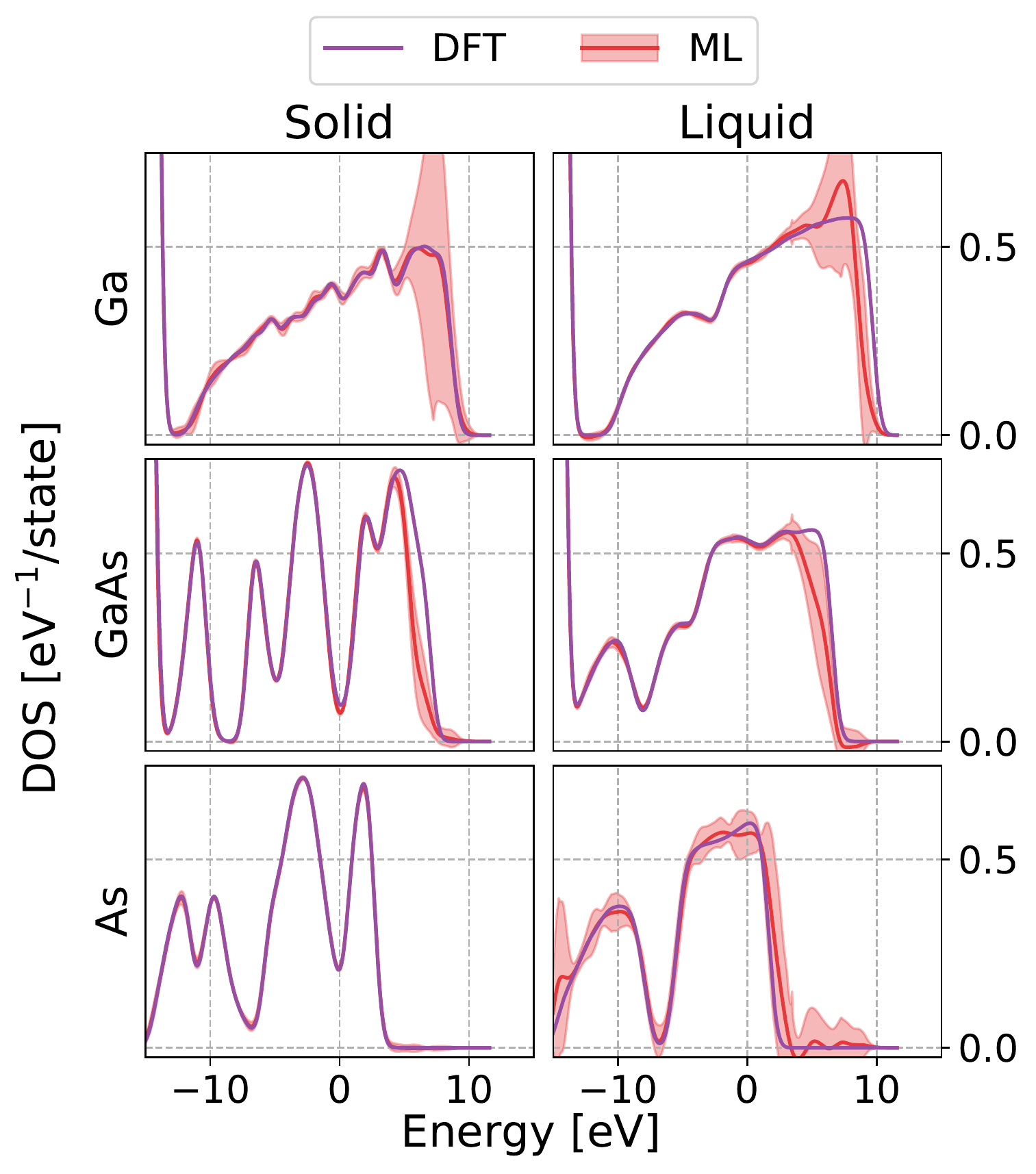}
  \caption{The electronic density of states of liquid and solid Ga, As, and GaAs as predicted by a committee of 16 built following Ref.~\citenum{imbalzano2020uncertainty}. All the curves are centered with respect to the Fermi energy, which represents the Energy=0 level.}
  \label{fig:GaAs_DOS}
\end{figure}

It is worth mentioning that the same transferability that is achieved for the potential also applies to other properties, such as those afforded by next-generation integrated ML models that also target predictions of the electronic structure of materials. 
As a proof of principle, we build a model of the DOS using the same protocol discussed in \cite{mahmoud2020learning}, using the Kohn-Sham eigenvalues from the same structures included in the training set for the potential, an additive decomposition of the DOS and a prediction of local contributions in terms of a multivariate Gaussian process regression and a description of atomic environments based on SOAP features\cite{bartok2013representing}, computed using the implementation in librascal\cite{musil2021rascal}. 
As shown in Fig. \ref{fig:GaAs_DOS}, this DOS model gives accurate predictions of the single-particle energy states across the entirety of the phase diagram. Even though the limitations of DFT-PBE (which is known to underestimate the band gap in GaAs) make this preliminary model of limited utility, future work may build on our results to incorporate electronic-structure information at a higher level of theory, such as it has been done for the Al$_x$Ga$_{1-x}$As system\cite{chan-vonl18prm}, providing a full description of the stability and properties of the \gaasx system.

\section{Conclusions}

We presented an accurate and flexible machine learning potential for the \gaasx binary system, that can be used to model technologically relevant phenomena and deepen our understanding of the atomistic processes that underlie them. We demonstrate an accuracy of this potential which is comparable to the electronic-structure reference calculations, while its low, and linear-scaling, computational cost will benefit the study of large-scale problems, such as the full nanowire-nanodroplet interface during the vapour-liquid-solid growth of GaAs nanowires, or the investigation of radiation damage in bulk GaAs. The transferability of this potential makes it also suitable to  study, with comparable accuracy, the behaviour of pure As and Ga. The latter, in particular, has become the subject of several works regarding the local structure at high pressure\cite{drewitt2020structural,mokshin2015short,li2017local} and the details of its liquid-liquid phase transition\cite{carvajal2009theoretical,li2016anomalous}. 
Both in the construction of the training set, and in the analysis of the simulation results we make use of the structure of the potential as a committee of multiple models\cite{musil2019fast}.
Their  standard deviation (after appropriate calibration) indicates the uncertainty of predictions for each configuration. This provides an online control of the quality of the trajectories that are generated and a reliable method to choose new training points to refine the fit. We also demonstrate how the committee model can be used to estimate the error for both static-lattice quantities and finite-temperature thermodynamic averages, \rev{assessing the error associated with the machine-learning approximation on the structure of the liquid phases and the values of the melting points.}\cite{imbalzano2020uncertainty} %

In order to provide compelling arguments for the quality of the potential, we have tested the predictions across many different scenarios. Some, such as the calculation of static lattice properties, are directly comparable to the DFT predictions. Others, such as finite-temperature static and dynamical quantities, cannot be directly computed from DFT and have been compared to their experimental counterparts. 
Overall, we have demonstrated that the potential is able to reproduce very well the reference data, with only minor deviations from the DFT calculations. We have also seen a good agreement with the experimental data on a number of properties. 
Finally, we have shown the flexibility and overall accuracy of the potential by exploring the finite-temperature properties of \gaasx, including challenging properties such as the melting point and the low-temperature thermophysical quantities, that require a quantum mechanical treatment of nuclear degrees of freedom. 
The binary phase diagram is in convincing agreement with the experimental one, although the shortcomings of the reference PBE DFT energetics are apparent in the underestimation of the melting point of all phases.
Higher-level-of-theory calculations, for instance using a hybrid DFT scheme, could increase the accuracy of this potential, as well as provide more reliable estimates of electronic excitations, and serve as an input for an integrated model providing both structural and functional properties of III-V semiconductors.

\section{Acknowledgements}

We thank Venkat Kapil for help in setting up the PIMD simulations and implementing the Np$_z$T barostat in i-PI, and Chiheb Ben Mahmoud for sharing code to compute a model of the electronic DOS. 
GI and MC acknowledge support by the NCCR MARVEL, funded by the Swiss National Science Foundation (SNSF).

\section{Supplementary Material}
The supplementary material contains additional data about bulk phases and point defects, and a comparison of the diffusion coefficients extrapolated using the viscosity or by finite-size scaling. 
The training data, the fitted potential, a chemiscope\cite{frau+20joss,chemiscope} visualization of the training set, and two examples of i-PI simulations using the potential are also included as a compressed archive.\cite{supplmat}

\end{document}